\documentclass{aa}  
\usepackage{graphicx}
\usepackage{txfonts}
\begin{document}
   \title{The circumburst environment of a FRED GRB: study of the
prompt emission and X-ray/optical afterglow of GRB~051111}


   \author{C. Guidorzi
          \inst{1}\fnmsep\thanks{\emph{Present address:} INAF - Osservatorio
	  Astronomico di Brera, via Bianchi 46, 23807 Merate (LC), Italy.
	  }
          \and
          A.~Gomboc\inst{1,2}
          \and
	  S.~Kobayashi\inst{1}
          \and
	  C.G.~Mundell\inst{1}
          \and
	  E.~Rol\inst{3}
          \and
	  M.F.~Bode\inst{1}
	  \and
	  D.~Carter\inst{1}
	  \and
	  V. La Parola\inst{4}
	  \and
	  A.~Melandri\inst{1}
	  \and
	  A.~Monfardini\inst{1,5}
	  \and
	  C.J.~Mottram\inst{1}
	  \and
	  P.T.~O'Brien\inst{3}
	  \and
	  K.L.~Page\inst{3}
	  \and
	  T.~Sakamoto\inst{6}
	  \and
	  R.J.~Smith\inst{1}
	  \and
	  I.A.~Steele\inst{1}
	  \and
	  N.R.~Tanvir\inst{3,7}
          }

   \offprints{C. Guidorzi - \email{cristiano.guidorzi@brera.inaf.it}}

   \institute{Astrophysics Research Institute, Liverpool John Moores University,
     Twelve Quays House, Egerton Wharf, Birkenhead, CH41 1LD, UK
   \and
   Faculty of Mathematics and Physics, University of Ljubljana, Jadranska 19,
   1000 Ljubljana, Slovenia
   \and
   Department of Physics \& Astronomy, University of Leicester,
   Leicester, LE1 7RH, UK
   \and
   INAF - Sezione di Palermo, Via U. La Malfa 153, 90146 Palermo, Italy
   \and
   ITC - IRST and INFN, Trento, via Sommarive, 18 38050 Povo (TN), Italy
   \and
   NASA Goddard Space Flight Center, Greenbelt, MD 20771, USA
   \and
   Centre for Astrophysics Research, University of Hertfordshire,
   Hatfield AL10 9AB, UK
   }

   \date{Received; accepted}

 
  \abstract
      {}
{We report a multi-wavelength analysis of the prompt emission and
early afterglow of GRB~051111 and discuss its properties in the
context of current fireball models.}
{The detection of GRB 051111 by the Burst Alert Telescope on-board
Swift triggered early BVRi' observations with the 2-m robotic Faulkes
Telescope North in Hawaii, as well as X-ray observations with the Swift
X-Ray Telescope.}
{The prompt $\gamma$-ray emission shows a
classical FRED profile. The optical afterglow light curves are fitted
with a broken power law, with ${\alpha}_1=0.35$ to ${\alpha}_2=1.35$ and
a break time around 12 minutes after the GRB. Although contemporaneous
X-ray observations were not taken, a power law connection between the
$\gamma$-ray tail of the FRED temporal profile and the late XRT flux
decay is feasible. Alternatively, if the X-ray afterglow tracks the
optical decay, this would represent one of the first GRBs for which
the canonical steep-shallow-normal decay typical of early X-ray
afterglows has been monitored optically. We present a detailed
analysis of the intrinsic extinction, elemental abundances and
spectral energy distribution.
From the absorption measured in the low X-ray band we find possible
evidence for an overabundance of some $\alpha$ elements such as
oxygen, [O/Zn]=$0.7\pm0.3$, or, alternatively,
for a significant presence of molecular gas.
The IR-to-X-ray Spectral Energy Distribution measured at 80 minutes
after the burst is consistent with the cooling break lying between
the optical and X-ray bands.
Extensive modelling of the intrinsic extinction suggests dust
with big grains or grey extinction profiles.
The early optical break is due either to an energy injection
episode or, less probably, to a
stratified wind environment for the circumburst medium.}
      {}


\titlerunning{The Early Multi--Colour Afterglow of GRB~051111}

   \maketitle
%

\section{Introduction}

The launch of the the {\em Swift} satellite in November 2004
(Gehrels et al.~\cite{Gehrels04}) ushered in a new era of rapid detection,
accurate localisation and observation of prompt $\gamma$-ray emission early
X-ray afterglows of Gamma-Ray Bursts (GRBs). Thus for the first time,
it became possible to explore the previously unknown behaviour of the
early X-ray afterglow from seconds to hours after the burst with
unprecedented sensitivity, as well as to trigger rapid GRB followup
observations from ground-based facilities working across the
electromagnetic spectrum. As a result, there is an increasing number
of GRBs with good early-time optical and X-ray coverage. Early
multi-band observations are crucial to discriminate between several
models; for example, the discovery that X-ray afterglow light curves
can exhibit a variety of complex properties such as flares and steep
early decays, has driven modifications to the standard fireball model
with late central engine activity suggested to explain flares and
continuous energy injection to account for the observed intermediate
shallow decay phase (e.g. Chincarini et al.
\cite{Chincarini05}; Nousek et al.~\cite{Nousek06}).

In addition, the combination of early optical and X-ray data sheds
light on the circumburst environment properties.
In particular, the presence of dust and its properties with
respect to gas can be studied through the optical extinction
vs. X-ray absorption. In this regard, in a seminal paper Galama
\& Wijers (\cite{Galama01})
found that the optical extinction is lower than expected from
the X-ray absorption measured in terms of equivalent hydrogen
column density; see also the comprehensive studies by Stratta
et al.~(\cite{Stratta04}) and Kann et al.~(\cite{Kann06}). 
For a number of GRBs at high redshift ($z\gtrsim4$) for
which an optical spectrum is available, a damped Lyman $\alpha$
absorption feature was observed.
These systems are called GRB Damped Lyman Systems
(GRB-DLAs) in analogy to the QSO-DLAs (Wolfe et al.~\cite{Wolfe05});
see Savaglio~(\cite{Savaglio06}) for an updated review of the properties
of both classes.
The properties that characterise the GRB-DLAs appear to be peculiar
and suggest a different ISM in the GRB hosts compared to that of the
Milky Way or Magellanic clouds, particularly for those GRBs at high redshift.
These peculiarities could be
ascribed to either the circumburst environment or to the
GRB host itself, or to the combination of the two.

A possible connection has been suggested between GRBs
whose $\gamma$-ray temporal profiles exhibit few pulses
and the nature of the progenitor as a SN (Bosnjak et al.~\cite{Bosnjak06},
Hakkila \& Giblin~\cite{Hakkila06}).
Despite the variety of temporal profiles of GRBs,
the so-called FRED GRBs (``Fast Rise Exponential Decay'')
seem to identify a subclass of GRBs (Fishman et al.~\cite{Fishman95}).
The multi-wavelength study of the early afterglow of this
kind of GRBs can be a further way to test this connection,
providing possible keys to the understanding of the variety
of the prompt light curves, still so poorly understood.

In this paper, we report the robotic detection and automatic
identification of the optical afterglow of a classical FRED detected
by {\em Swift}, GRB~051111, using the 2-m Faulkes Telescope
North (FTN) located in Maui, Hawaii, from 3.9 to 168 minutes after
the GRB trigger time. We also report the {\em Swift}/BAT and XRT
observations of the prompt $\gamma$-ray event and of the X-ray
afterglow, respectively.
The redshift of this burst is known spectroscopically:
$z=1.55$ (Hill et al.~\cite{Hill05}).
The paper is structured as follows. In Sec.~\ref{sec:obs} observations
are reported. The results are presented in Sec.~\ref{sec:res}.
In Sec.~\ref{sec:SED} we discuss the Spectral Energy Distribution
(SED) at $t=80$ minutes after the burst, when we have the best
clustering of observations from IR to X-rays. In particular,
we derive clues on the circumburst environment properties by
comparing the intrinsic optical extinction and the X-ray
absorption. In Sec.~\ref{sec:disc} we discuss some possible
interpretations of the overall picture.

\section{Observations}
\label{sec:obs}
\subsection{{\em Swift}/BAT Observations}
\label{sec:obs_BAT}
On 2005 November 11 at 05:59:41 UT
GRB~051111 triggered {\it Swift}/BAT, which
automatically on board provided the location
of the burst with an error radius of 3 arcmin (90\% CL; Sakamoto et
al.~\cite{Sakamoto05}). During the prompt event, the spacecraft did
not slew because of the angular proximity of the GRB to the Moon.
Data have been reduced following the standard BAT pipeline
(Krimm et al.~\cite{Krimm04}) with the HEASARC {\em heasoft} package (v. 6.0.4):
we created a detector plane image and a merged quality mask
and finally we added a mask weight column to the event file
using the direction of the burst (see Section~\ref{sec:obs_opt}).
Mask-weighted light curves and time-resolved spectra
were extracted.
Spectra have been corrected for systematics with the tool
{\em batphasyserr} as recommended by the BAT team
\footnote{http://heasarc.gsfc.nasa.gov/docs/swift/analysis/bat\_digest.html .}.

\subsection{{\em Swift}/XRT Observations}
\label{sec:obs_XRT}
XRT began observing the afterglow of GRB~051111 from 95~minutes
to about 900~minutes after the burst. We consider here only
the data in Photon Counting (PC) mode.
Data have been processed following the standard XRT pipeline
with the ftool {\em xrtpipeline} (v. 0.9.9) using standard
screening criteria. We extracted light curves and spectra
from a 20-pixel radius circle centred on the optical afterglow.
We found that pile up was negligible, following Vaughan et al.~(\cite{Vaughan06}).
Background was determined from four 50-pixel radius circles
with no sources. Spectral analysis was done adopting either
empirical and physical ancillary files. We adopted the
response function of the latest distribution of the HEASARC
calibration database (CALDB~2.3).

\subsection{Faulkes Telescope North Observations}
\label{sec:obs_opt}
The optical afterglow was discovered
by the ROTSE-IIIb telescope 27~s after the burst
within the BAT error circle and with an unfiltered magnitude of 13.0
(Rujopakarn et al.~\cite{ROTSE05}).

The optical afterglow was soon confirmed by the FTN
which robotically followed up GRB~051111 at 3.87~min after
the burst trigger time (3.5~min after the GRB notice time).
The optical transient (OT) candidate
was identified automatically and independently of ROTSE-IIIb as a fading
uncatalogued source with initial magnitude $R\sim14.9$
(Mundell~et~al.~\cite{Mundell05}). 
The automatic identification of the OT with the highest confidence
level (1.0) by the GRB pipeline ``LT-TRAP'' (Guidorzi et al.~\cite{Guidorzi06})
triggered a multi-colour imaging sequence: however, after the first
3x10~s $R$ images, followed by 3x10~s in $B$, $V$ and
Sloan $i'$, the telescope unexpectedly stopped observing
due to technical problems. A manual intervention restored the
follow-up observation program at $\sim$36~min after the burst.
The sequence of filters used reflects the strategy described
by Guidorzi et~al.~(\cite{Guidorzi06}) for the multicolour image mode (MCIM)
for the FTN.
A late follow-up observation with FTN was performed from
100 to 168 minutes after the burst in the $BVRi'$ filters as
part of the RoboNet
project\footnote{http://www.astro.livjm.ac.uk/RoboNet/}.
Table~\ref{tab:obs} summarises the observations with FTN.

\begin{table}
\caption{Optical photometry for GRB~051111 with the Faulkes Telescope North}
\label{tab:obs}
\centering
\begin{tabular}{lrrrcl}
\hline\hline
Filter & Start$^{a}$ & End$^{a}$ & Exp. & Mag. & Comment$^{b}$\\
       & (min)       & (min)     & (s)      &      &\\
\hline
Bessell-R &    3.87 &  4.04 &      10 & $14.92\pm0.03$ & DM\\
Bessell-R &    4.23 &  4.40 &      10 & $14.98\pm0.03$ & DM\\
Bessell-R &    4.58 &  4.75 &      10 & $15.03\pm0.03$ & DM\\
Bessell-B &    5.88 &  6.05 &      10 & $16.32\pm0.04$ & MCS\\
Bessell-V &    6.72 &  6.89 &      10 & $15.87\pm0.03$ & MCS\\
SDSS-I    &    7.63 &  7.80 &      10 & $15.30\pm0.02$ & MCS\\
Bessell-R &    36.1 &  37.0 &    3x10 & $16.99\pm0.09$ & DM\\
Bessell-B &   38.18 & 38.35 &      10 & $18.20\pm0.14$ & MCS\\
Bessell-V &   39.03 & 39.20 &      10 & $17.82\pm0.14$ & MCS\\
SDSS-I    &   39.95 & 40.12 &      10 & $17.07\pm0.07$ & MCS\\
Bessell-B &    41.6 &  42.1 &      30 & $18.30\pm0.08$ & MCS\\
Bessell-R &    42.9 &  43.4 &      30 & $17.22\pm0.07$ & MCS\\
SDSS-I    &    44.2 &  44.7 &      30 & $17.19\pm0.04$ & MCS\\
Bessell-B &    45.4 &  46.4 &      60 & $18.48\pm0.08$ & MCS\\
Bessell-R &    47.3 &  48.3 &      60 & $17.33\pm0.05$ & MCS\\
SDSS-I    &    49.1 &  50.1 &      60 & $17.40\pm0.03$ & MCS\\
Bessell-B &    50.8 &  52.8 &     120 & $18.41\pm0.05$ & MCS\\
Bessell-R &    53.6 &  55.6 &     120 & $17.47\pm0.04$ & MCS\\
SDSS-I    &    56.3 &  69.6 & 120+180 & $17.56\pm0.03$ & MCS\\
Bessell-B &    59.1 &  62.1 &     180 & $18.80\pm0.05$ & MCS\\
Bessell-R &    62.8 &  65.8 &     180 & $17.79\pm0.04$ & MCS\\
Bessell-B &    70.4 &  72.4 &     120 & $18.81\pm0.08$ & MCS\\
Bessell-R &    73.2 &  75.2 &     120 & $17.78\pm0.05$ & MCS\\
SDSS-I    &    76.0 &  89.4 & 120+180 & $17.97\pm0.04$ & MCS\\
Bessell-R &    82.6 &  85.6 &     180 & $18.13\pm0.06$ & MCS\\
Bessell-B &    90.1 &  94.1 &     240 & $19.23\pm0.09$ & MCS\\
Bessell-R &    95.0 &  99.0 &     240 & $18.25\pm0.06$ & MCS\\
Bessell-R &   100.6 & 108.4 &   3x150 & $18.34\pm0.06$ & LFU\\
Bessell-R &   108.6 & 111.1 &     150 & $18.44\pm0.08$ & LFU\\
Bessell-V &   112.5 & 123.1 &   4x150 & $18.95\pm0.10$ & LFU\\
Bessell-B &   124.4 & 135.0 &   4x150 & $19.75\pm0.14$ & LFU\\
SDSS-I    &   157.0 & 167.5 &   4x150 & $18.66\pm0.09$ & LFU\\
\hline
\end{tabular}
\flushleft
$^{a}$ This corresponds to the time delay with
respect to the GRB trigger time, $t_{\rm trig}=21581.312$~SOD UT.\\
$^{b}$ DM, MCS and LFU stand for ``Detection Mode'', ``Multi-Colour Sequence''
and ``Late Follow Up'', respectively.
\end{table}

Magnitudes in $BVR$ have been calibrated using Landolt standard
field stars (Landolt~\cite{Landolt92}). The zero points derived from observations
of standards before and after the GRB observations were consistent.
Magnitudes in $i'$ have been calibrated using the same Landolt
standard field stars, after converting from Bessell $RI$ values
to SDSS $r'i'$ assuming power-law spectra.
Magnitudes have been corrected for the airmass and Galactic extinction.
The latter was estimated from the $E_{B-V}=0.159$ derived from the 
extinction maps by Schlegel et al.~(\cite{Schlegel98}),
and $A_V=R_V\cdot E_{B-V}=0.49$, with $R_V=3.1$.
We evaluated the extinction in the other filters
following the parametrisation by Cardelli et al.~(\cite{Cardelli89}):
$A_B=0.64$, $A_{R}=0.40$ and $A_{i'}=0.32$.
Magnitudes have been converted into flux densities $F_{\nu}$ (mJy)
following Bessell~(\cite{Bessell79}) and Fukugita et al.~(\cite{Fukugita96})
for the $BVR$ and $i'$ pass-bands, respectively.

\section{Results}
\label{sec:res}
All the reported parameter uncertainties are given at
1-$\sigma$ confidence level (CL), unless stated otherwise.

\subsection{The $\gamma$-ray prompt emission}
\label{sec:res_BAT}
\begin{figure}
  \centering
\resizebox{\hsize}{!}{\includegraphics{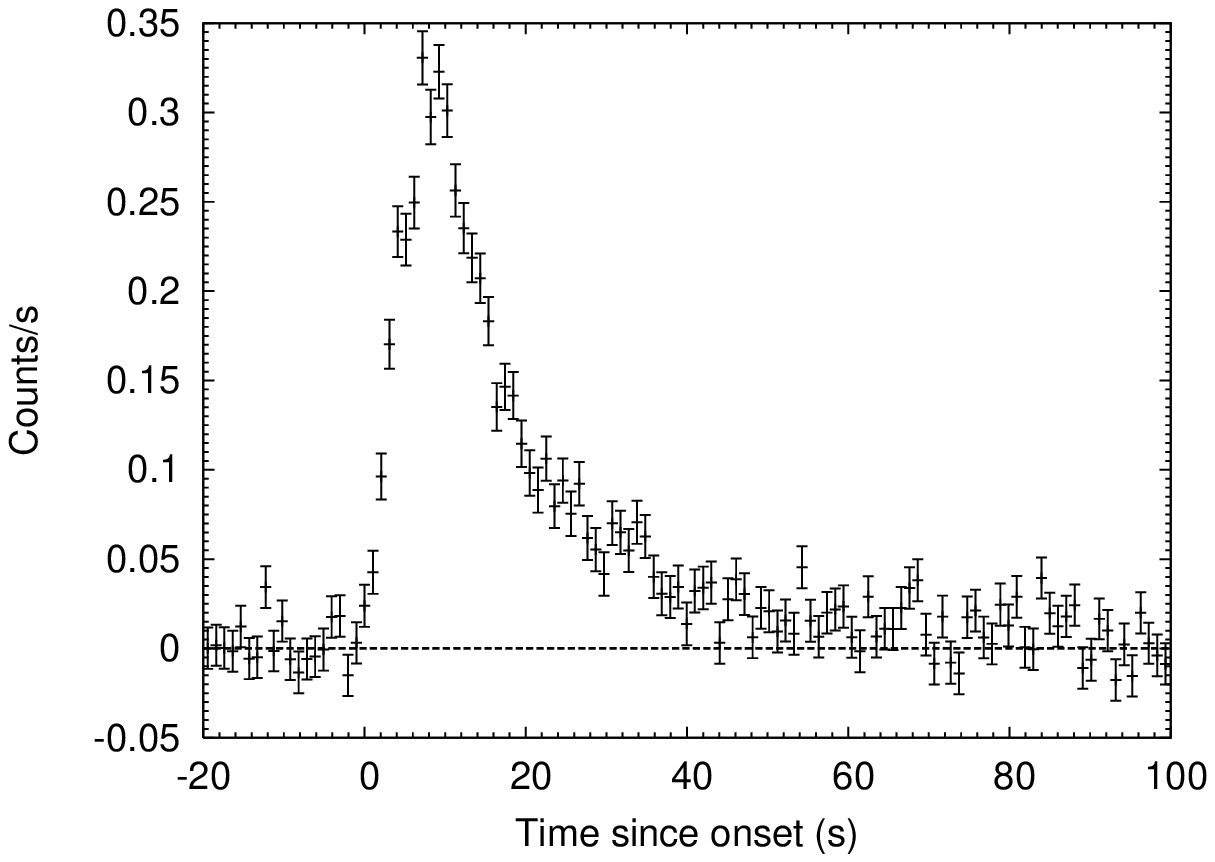}}
\resizebox{\hsize}{!}{\includegraphics{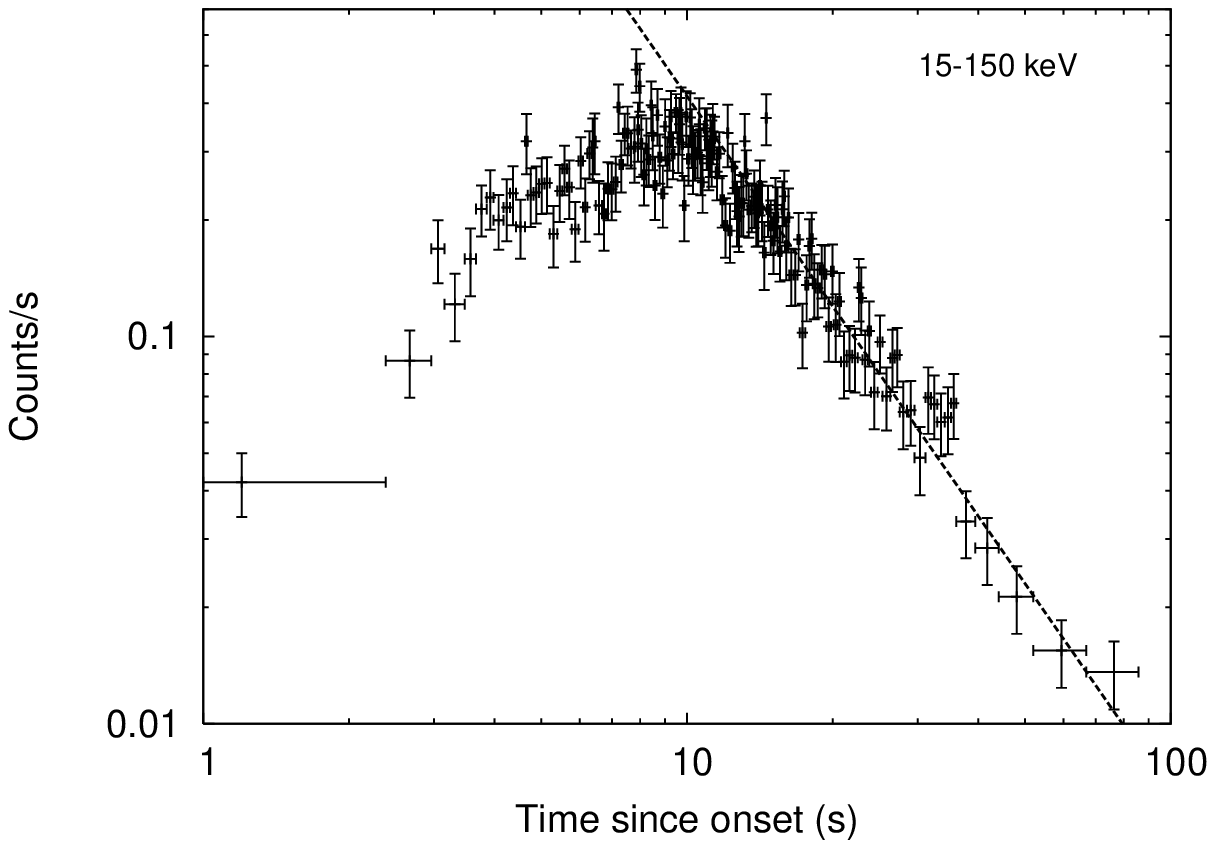}}
\caption{{\em Top Panel}: {\em Swift}/BAT 15--150~keV mask-weighted light curve
of GRB~051111. The burst onset time corresponds to 21573.312
seconds of day (corresponding to 8 s earlier than the trigger time).
{\em Bottom Panel}: the same light curve in log scale. Dashed line
shows the best-fit power-law applied to $t>18~s$ data:
$\alpha_\gamma=1.8\pm0.2$ ($F_\gamma\sim t^{-\alpha_\gamma}$).}
\label{fig:BAT_lin_lc}
\end{figure}

\begin{figure}
  \centering
\resizebox{\hsize}{!}{\includegraphics{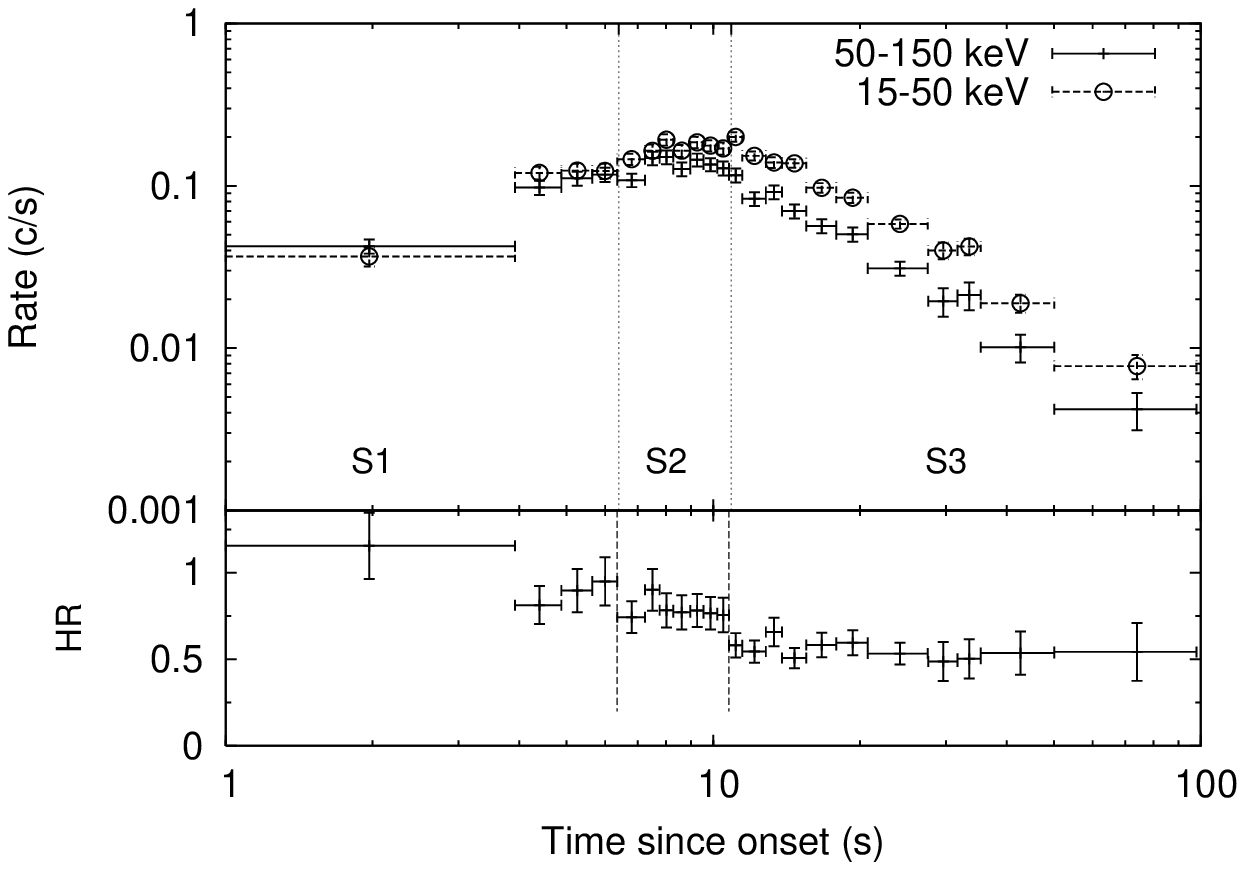}}
\caption{{\em Top Panel}: BAT light profiles of the mask-weighted
light curve in two energy channels: 15--50~keV and 50--150~keV.
{\em Bottom Panel}: the hardness ratio between the two energy channels
exhibits a typical hard-to-soft evolution.
The three temporal intervals correspond to the three different energy
spectra discussed in the text.}
\label{fig:BAT_hr}
\end{figure}

\begin{figure}
  \centering
\resizebox{\hsize}{!}{\includegraphics{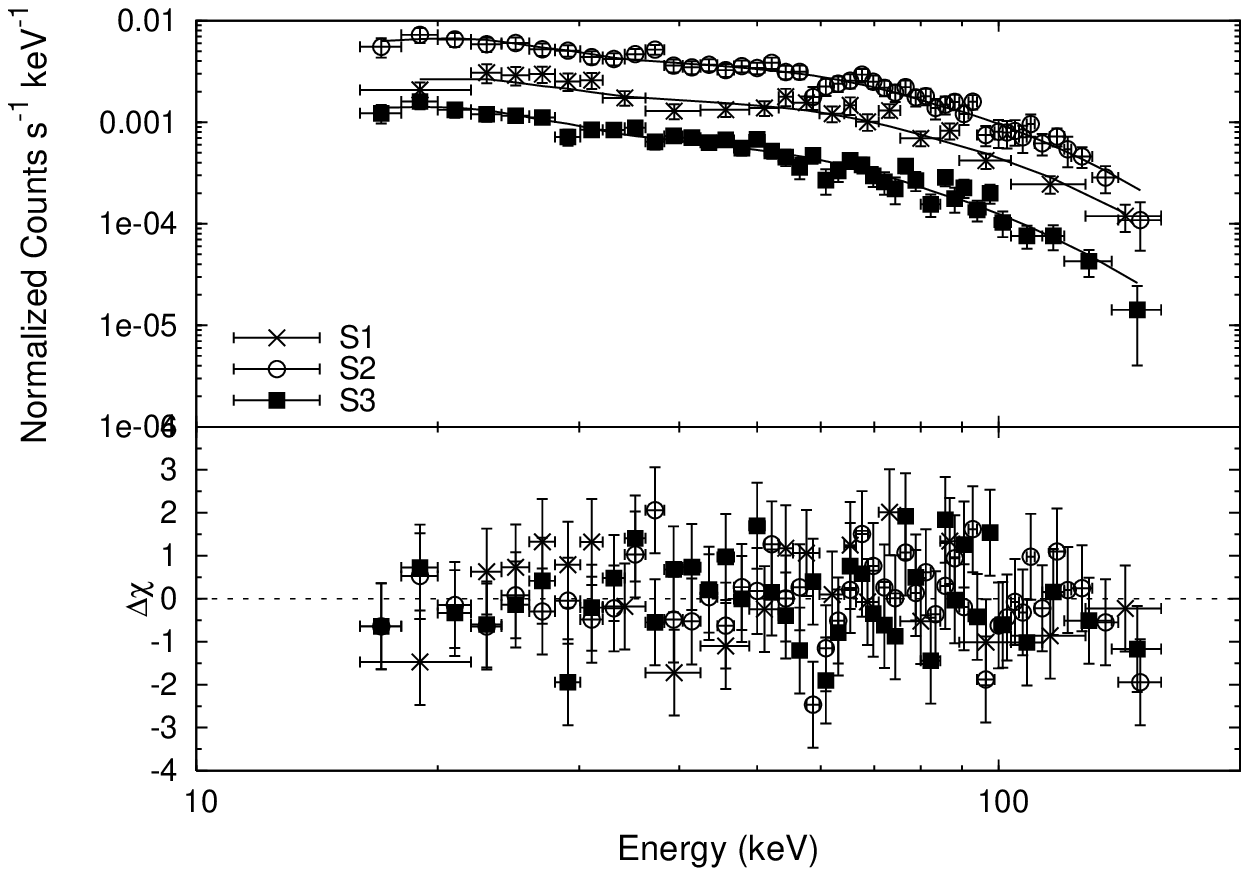}}
\caption{15--150~keV BAT spectra extracted in the three temporal
intervals shown in Fig.~\ref{fig:BAT_hr}.}
\label{fig:BAT_all_spectra}
\end{figure}

The 15--150~keV temporal profile is that of a typical FRED
and the mask-weighted light
curve shows significant emission above the background up
to about 90~s after the burst onset time (Fig.~\ref{fig:BAT_lin_lc}).
The onset of this burst was 8~s before the trigger time, which is
21581.312~SOD UT, and is assumed as zero time hereafter.
The spectral evolution is typical, from hard to soft,
as shown by the hardness ratio temporal evolution
between the 15--50~keV and 50--150~keV
(Fig.~\ref{fig:BAT_hr}).
The fit of the $\gamma$-ray tail alone with a power law
for $t>18$~s gives an acceptable result, with a slope of
$\alpha_\gamma=1.8\pm0.2$ ($F_\gamma\propto t^{-\alpha_\gamma}$).
The average spectrum is fitted with a power law with photon
index $\Gamma_{\gamma}=1.33\pm 0.04$, where
$dN/dE\propto E^{-\Gamma_{\gamma}}$ is the photon spectrum.
This is in agreement with the value of $1.5\pm0.3$ measured by the {\em Suzaku}
Wideband All-sky Monitor (WAM) in the 100--700~keV band (Yamaoka et al.~\cite{Yamaoka05}).
The 15--150~keV fluence is $(4.2\pm0.2)\times 10^{-6}$~erg~cm$^{-2}$.
The 15--150~keV peak flux is $(2.2\pm0.1)\times 10^{-7}$~erg~cm$^{-2}$~s$^{-1}$.
\begin{table*}
\caption{{\em Swift}/BAT 15--150~keV of the $\gamma$-ray prompt
emission: best-fit results obtained with a power-law model.}
\label{tab:BAT_spec}
\centering
\begin{tabular}{lccccr}
\hline\hline
Spectrum & Start Time$^{a}$ & End Time$^{a}$ & $\Gamma_{\gamma}$ & Average Flux & $\chi^2/{\rm dof}$\\
         & (s)              & (s)            &                   & ($10^{-7}$~erg~cm$^{-2}$~s$^{-1}$) & \\
\hline
S1  & 0.02 & 14.35   & $1.07\pm 0.07$ & $0.97\pm0.04$ & 23.5/19\\
S2  & 14.35 & 18.77 & $1.15\pm 0.05$ & $2.18\pm0.06$ & 36.1/46\\
S3  & 18.77 & 98.02 & $1.44\pm 0.05$ & $0.33\pm0.01$ & 37.6/39\\
whole & 0.02 & 98.02 & $1.33\pm 0.04$ & $0.47\pm0.01$ & 39.5/47\\
\hline
\end{tabular}\\
\flushleft
$^{a}$ Time with respect to the GRB onset: $t_{\rm onset}=21573.312$~SOD UT.
\end{table*}
We extracted the 15--150~keV spectrum in three different temporal
slices, named ``S1'', ``S2'' and ``S3'', corresponding to the rise,
around the peak, and the tail, respectively (see Fig.~\ref{fig:BAT_hr}).
The choice of these spectra was driven by the light curve and the
hardness ratio evolution displayed in Fig.~\ref{fig:BAT_hr}: each
boundary (between S1 and S2, and between S2 and S3) seems to mark a
change in the hardness ratio. In addition, the fact that each
slice identifies a physically different portion of the light curve,
i.e. rise, peak or tail, which is probably related to correspondingly
different regimes, led us to attempt to disentangle possible distinctive features.
Each spectrum has been binned in order to have at least 5-$\sigma$ signal
in each grouped channel.
We fitted each spectrum with a power law.
The hard-to-soft spectral evolution is confirmed by the photon index
$\Gamma_{\gamma}$ increasing with time.
The best-fit spectral results are reported in Table~\ref{tab:BAT_spec}
and shown in Fig.~\ref{fig:BAT_all_spectra}.
%

\subsection{The X-ray afterglow}
\label{sec:res_XRT}
\begin{figure*}
  \centering
\resizebox{\hsize}{!}{\includegraphics{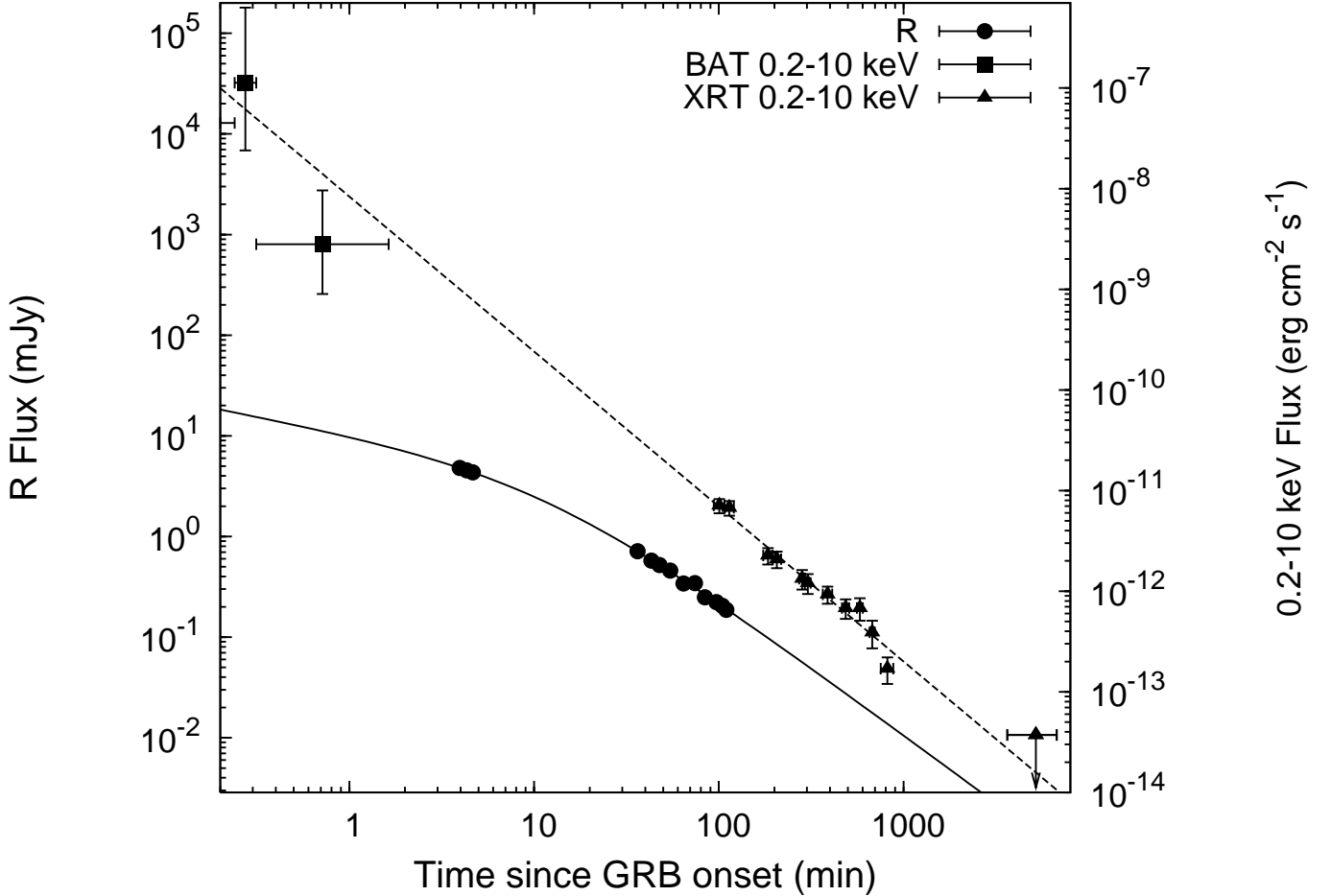}}
\caption{$R$ and X-ray light curves of GRB~051111 (observer frame).
Solid line shows the best smooth broken power-law fit of $R$ points
(see text). 0.2--10~keV flux derived from BAT (squares) and XRT
(triangles) data is also shown.
Dashed line shows the
best fit power-law connecting the tail of BAT data up to the XRT data.
The last XRT point is a 3-$\sigma$ upper limit.}
\label{fig:all_lc_x}
\end{figure*}

The 0.2--10~keV light curve, from 95~minutes to $\sim$900~minutes
after the burst (Fig.~\ref{fig:all_lc_x}), can be
fit with a single power law, $F_{\rm x}(t)\sim t^{-\alpha_{\rm x}}$,
with $\alpha_{\rm x}=1.56\pm0.08$  ($\chi^2/{\rm dof}=11.07/9$).

Alternatively, we found that the X-ray light curve may also be
described by three power-law segments: $\alpha_{1,{\rm x}}=1.65\pm0.12$,
$\alpha_{2,{\rm x}}=0.9\pm0.6$, $\alpha_{3,{\rm x}}=4.1\pm2.2$ (this is based
only on two points), with break times at $t_{1,{\rm x}}=300\pm110$~min and
$t_{2,{\rm x}}=600\pm1100$~min ($\chi^2/{\rm dof}=14.5/20$).
However if we compare this description with that of the majority
of the {\em Swift} bursts' light curves exhibiting the canonical
steep-shallow-normal decay, for which it is typically $t_{1,{\rm x}}<20$~min
(see Nousek et al.~\cite{Nousek06}, O'Brien et al.~\cite{OBrien06}),
in the case of the GRB~051111 the first break would occur unusually late,
implying a correspondingly unusually short shallow-decay phase.

The 0.2--10~keV spectrum has been extracted in three temporal
intervals: from 95 to 120 minutes, from 174 to 313 minutes and
from 365 to 700 minutes after the GRB.
We rebinned the energy channels of the earliest spectrum in order
to have 5$\sigma$ significant counts for each channel.
We rebinned the other spectra applying the same binning scheme
to allow a better comparison between the different spectra.
\begin{figure}
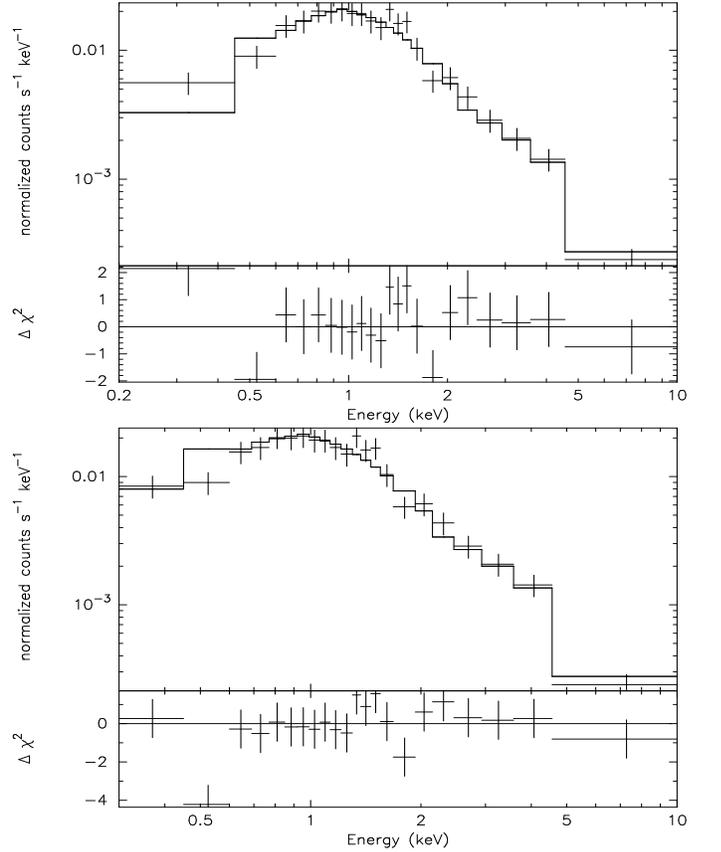

\centering
\resizebox{\hsize}{!}{\includegraphics[width=5.0cm,height=8cm,angle=270]{grb051111_xrt_spec_whole_5sig.ps}}
\resizebox{\hsize}{!}{\includegraphics[width=5.0cm,height=8cm,angle=270]{grb051111_xrt_spec_whole_5sig_03keV_ignore05keV.ps}}
\caption{{\em Top Panel}: 0.2--10~keV XRT spectrum fit with an absorbed power law in excess
to Galactic absorption (see text). {\em Bottom Panel}: 0.3--10~keV XRT spectrum fit with the same
model, ignoring the channel between 0.43 and 0.6~keV
(whose residual with respect to the fit is shown).}
\label{fig:XRT_spec}
\end{figure}
We fitted each spectrum
adopting either empirical or physical ancillary files
(Capalbi et al.~\cite{Capalbi05}):
in each case, we find consistent results within errors (see
Table~\ref{tab:XRT_spec}). Hereafter, we consider the results
obtained with empirical files only. We find no evidence for
significant spectral changes: the spectrum can be fit with
a power law with intrinsic photoelectric absorption in addition
to the Galactic one, $N_{\rm H}^{\rm (Gal)}=5.82\times10^{20}$~cm$^{-2}$
(Kalberla~et~al.~\cite{Kalberla05}).
Hereafter, $N_{\rm H}$ is the intrinsic rest-frame equivalent
hydrogen column density responsible for the soft X-ray absorption,
as evaluated with the XSPEC model {\sc zwabs}.
These results are consistent with those reported by La Parola et al.~(\cite{LaParola05}).
From the total spectrum, we find significant absorption:
$N_{\rm H}=8.0_{-2.6}^{+3.3}\times10^{21}$~cm$^{-2}$ (90\% CL) and a
photon index of $\Gamma_{\rm x}=2.15\pm0.15$ (corresponding to the
average of the values obtained assuming empirical and physical
ancillary files of $2.1\pm0.1$ and $2.2\pm0.1$, respectively).
However, we point out that this $N_{\rm H}$ measurement assumes a solar
metallicity.
Alternatively, if we impose a 1/10 solar metallicity as assumed by
Penprase et al.~(\cite{Penprase06}),
we find $N_{\rm H}=7.1_{-2.4}^{+3.2}\times10^{22}$~cm$^{-2}$ (90\% CL),
i.e. much higher than the estimate of $7.9\times10^{21}$~cm$^{-2}$
inferred by the Zn II column density measured with the Keck spectrum
under the 1/10 solar metallicity assumption (Penprase et al.~\cite{Penprase06}).
We tested the robustness of the $N_{\rm H}$ measurement in three different cases.
We focused on the total spectrum, given the lack of significant spectral evolution.
\begin{enumerate}
\item We rebinned the 0.2--10~keV spectrum by imposing a threshold of
5 $\sigma$ on the significance of each grouped energy channel;
\item we considered the 0.3--10~keV energy band (recommended in the April 2006 release
of the XRT CALDB \footnote{http://swift.gsfc.nasa.gov/docs/heasarc/caldb/swift/docs/xrt/index.html});
\item we considered the 0.3--10~keV and ignored the grouped channel around 0.5~keV (from
0.43~keV to 0.6~keV) to exclude the possible effect of an instrumental absorption feature
due to Oxygen, as warned in the case of bright sources.
\end{enumerate}
The results are consistent within uncertainties with those reported above.
Figure~\ref{fig:XRT_spec} shows the cases corresponding to item 1 (top panel) and
to item 3 (bottom panel), respectively.
The value derived for the $N_{\rm H}$ ({\sc zwabs pow} model)
decreases to $5.1_{-2.1}^{+2.9}\times10^{21}$~cm$^{-2}$ (90\% CL)
when the channel around 0.5~keV is ignored in the fit. This is little smaller than the value
reported above, although still consistent with it and still in excess of the Galactic value.
All the results are reported in Table~\ref{tab:XRT_spec}.
\begin{table*}
\caption{{\em Swift}/XRT 0.2--10~keV of the X-ray afterglow:
best-fit results obtained with an absorbed (intrinsic, in addition
to the Galactic, fixed to $N_{\rm H}^{\rm (Gal)}=5.82\times10^{20}$~cm$^{-2}$)
power-law model. $N_{\rm H}$ is
expressed in the GRB rest-frame. We adopted the solar abundances
by Asplund et al.~(\cite{Asplund05}). Parameters uncertainties are reported at
90\% CL.}
\label{tab:XRT_spec}
\centering
\begin{tabular}{lccccccr}
\hline\hline
XSPEC & Start Time$^{a}$ & End Time$^{a}$ & $N_{\rm H}$ & $Z/Z_{\odot}$ & $\Gamma_{\rm x}$ &
Average Flux & $\chi^2/{\rm dof}$\\
Model   & (min) & (min) & ($10^{21}$~cm$^{-2}$) &  &   & ($10^{-12}$~erg~cm$^{-2}$~s$^{-1}$) & \\
\hline
{\sc zwabs pow}$^{d}$  & 95 & 120  & $10_{-4}^{+7}$ & [1] & $2.3_{-0.3}^{+0.3}$ & 6.4 & 7.8/7\\
{\sc zwabs pow}$^{d}$  & 174 & 313 & $4.3_{-3.1}^{+4.3}$ & [1] & $2.2_{-0.3}^{+0.3}$ & 1.6 & 2.3/7\\
{\sc zwabs pow}$^{d}$  & 365 & 700 & $10_{-6}^{+10}$ & [1] & $2.0_{-0.2}^{+0.4}$ & 0.7 & 6.6/7\\
{\sc zwabs pow}$^{d}$  & 95  & 700 & $8.0_{-2.6}^{+3.3}$ & [1] & $2.2_{-0.2}^{+0.2}$ & 1.3 & 8.13/7\\ 
{\sc zwabs pow}$^{d}$  & 95  & 700 & $8.5_{-3.0}^{+3.7}$ & [1] & $2.2_{-0.2}^{+0.2}$ & 1.3 & 20.0/19$^{f}$\\ 
{\sc zwabs pow}$^{d}$  & 95  & 700 & $8.6_{-3.0}^{+3.8}$ & [1] & $2.2_{-0.2}^{+0.2}$ & 1.3 & 20.5/19$^{f,g}$\\ 
{\sc zwabs pow}$^{d}$  & 95  & 700 & $5.1_{-2.1}^{+2.9}$ & [1] & $2.2_{-0.2}^{+0.1}$ & 1.3 & 12.0/18$^{f,h}$\\ 
{\sc zvarabs pow}$^{d}$ & 95  & 700 & $71_{-24}^{+32}$ & [0.1]$^{b}$ & $2.2_{-0.2}^{+0.1}$ & 1.3 & 8.7/7\\
{\sc zvarabs pow}$^{d}$ & 95  & 700 & [7.94]$^{c}$ & $1.5_{-0.7}^{+0.7}$ & $2.2_{-0.2}^{+0.2}$ & 1.3 & 8.0/7\\
{\sc zvarabs pow}$^{d}$ & 95  & 700 & $45_{-15}^{+20}$ & [0.2] & $2.2_{-0.2}^{+0.2}$ & 1.3 & 8.4/7\\
{\sc zvarabs pow}$^{d}$ & 95  & 700 & $33_{-11}^{+15}$ & [0.3] & $2.2_{-0.2}^{+0.2}$ & 1.3 & 8.3/7\\
                                   &     &     &                         &                     &                     &     &\\
{\sc zwabs pow}$^{e}$  & 95 & 120  & $9_{-4}^{+5}$ & [1] & $2.2_{-0.3}^{+0.3}$ & 4.8 & 8.6/7\\
{\sc zwabs pow}$^{e}$  & 174 & 313 & $3.5_{-2.9}^{+4.0}$ & [1] & $2.1_{-0.2}^{+0.4}$ & 1.2 & 3.1/7\\
{\sc zwabs pow}$^{e}$  & 365 & 700 & $8_{-5}^{+7}$ & [1] & $1.9_{-0.3}^{+0.4}$ & 0.6 & 5.6/7\\
{\sc zwabs pow}$^{e}$  & 95  & 700 & $6.6_{-2.3}^{+2.9}$ & [1] & $2.1_{-0.1}^{+0.2}$ & 1.0 & 8.8/7\\ 
{\sc zwabs pow}$^{e}$  & 95  & 700 & $6.4_{-2.4}^{+3.3}$ & [1] & $2.1_{-0.2}^{+0.2}$ & 1.0 & 24.7/19$^{f}$\\ 
{\sc zwabs pow}$^{e}$  & 95  & 700 & $6.4_{-2.6}^{+3.5}$ & [1] & $2.1_{-0.2}^{+0.2}$ & 1.0 & 25.4/19$^{f,g}$\\ 
{\sc zwabs pow}$^{e}$  & 95  & 700 & $4.1_{-1.9}^{+2.5}$ & [1] & $2.1_{-0.2}^{+0.2}$ & 1.0 & 16.2/18$^{f,h}$\\ 
{\sc zvarabs pow}$^{e}$ & 95  & 700 & $59_{-22}^{+28}$ & [0.1]$^{b}$ & $2.1_{-0.2}^{+0.2}$ & 1.0 & 8.8/7\\
{\sc zvarabs pow}$^{e}$ & 95  & 700 & [7.94]$^{c}$ & $1.2_{-0.4}^{+0.6}$ & $2.1_{-0.1}^{+0.2}$ & 1.0 & 8.8/7\\
{\sc zvarabs pow}$^{e}$ & 95  & 700 & $38_{-14}^{+17}$ & [0.2] & $2.1_{-0.1}^{+0.2}$ & 1.0 & 8.8/7\\
{\sc zvarabs pow}$^{e}$ & 95  & 700 & $28_{-10}^{+12}$ & [0.3] & $2.1_{-0.1}^{+0.2}$ & 1.0 & 8.8/7\\
\hline
\end{tabular}\\
\flushleft
$^{a}$ Time with respect to the GRB onset: $t_{\rm onset}=21581.312$~SOD UT.\\
$^{b}$ Penprase et al.~(\cite{Penprase06}) assumed 1/10 solar metallicity.\\
$^{c}$ Penprase et al.~(\cite{Penprase06}) derived $N_{\rm H}=7.94\times 10^{21}$~cm$^{-2}$.\\
$^{d}$ Empirical ancillary files adopted.\\
$^{e}$ Physical ancillary files adopted.\\
$^{f}$ Different grouping adopted: each grouped channel is more significant than 5 $\sigma$.\\
$^{g}$ Energy band considered: 0.3--10~keV.\\
$^{h}$ Energy band considered: 0.3--10~keV and channel around 0.5~keV ignored.\\
\end{table*}
%

\subsection{BAT/XRT joint light curve}
\label{sec:res_BATXRT}
Following O'Brien et al.~(\cite{OBrien06}),
for each of the three time intervals considered in Sec.~\ref{sec:res_BAT},
we extrapolated the 15--150~keV fluxes into unabsorbed fluxes in the
0.2--10~keV band assuming the mean of the corresponding BAT photon
index ($\Gamma_{\gamma}=1.07, 1.15, 1.44$, respectively) and that measured
with the XRT ($\Gamma_{\rm x}=2.15\pm0.15$).
As it is not known where the break energy lies, uncertainties were evaluated
by assuming as photon index the two boundary values, i.e. either the
corresponding BAT or the XRT photon index.
The resulting 0.2--10~keV light curve is shown in Fig.~\ref{fig:all_lc_x}.
We fitted the light curve from 10~s to 1000~min after the GRB onset
with a simple power law: $\alpha_{{\rm x},\gamma}=1.54\pm0.10$
($\chi^2/{\rm dof}=6.2/11$), consistent with the power-law
decay shown in late X-rays alone.

\subsection{The optical afterglow}
\label{sec:res_opt}
\begin{figure*}
  \centering
\resizebox{\hsize}{!}{\includegraphics{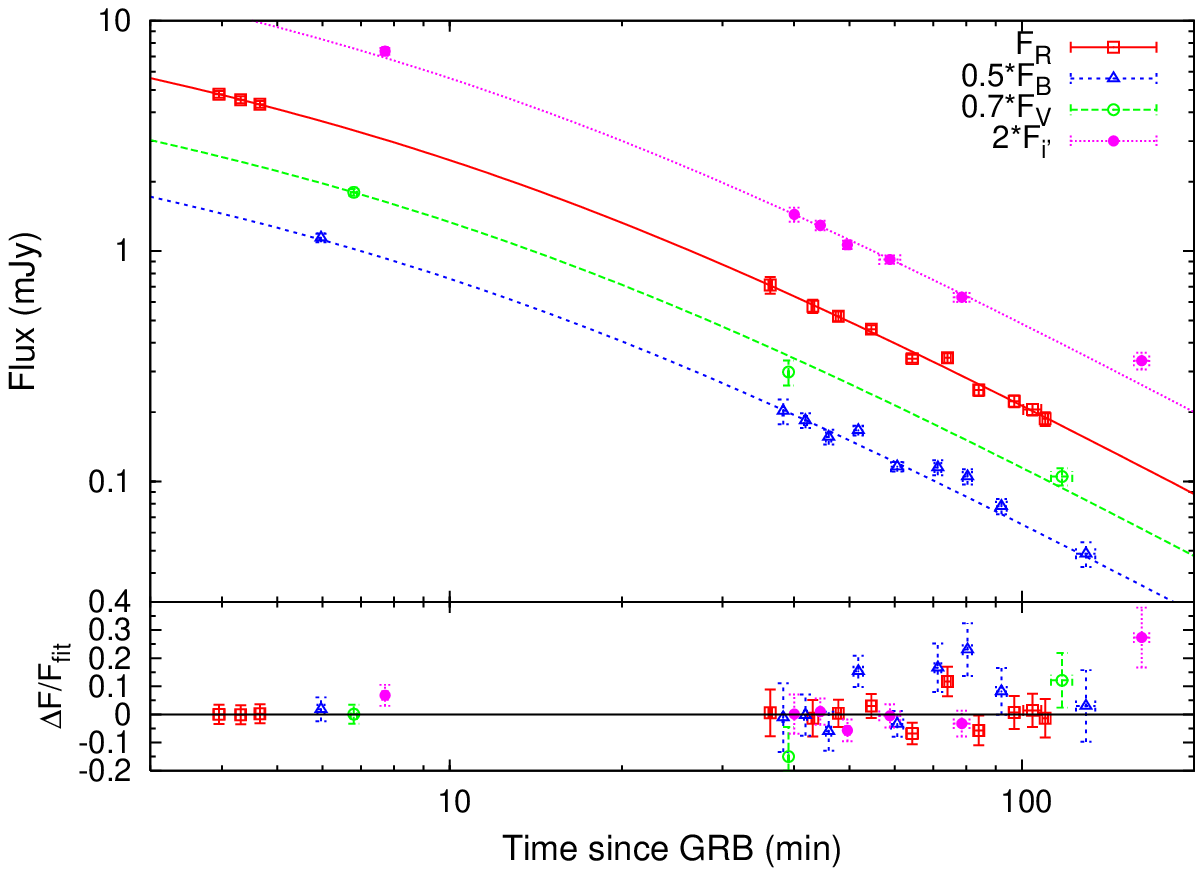}}
\caption{{\em Top panel}:
$BVRi'$ multi-colour light curve of GRB~051111 acquired by FTN
(observer frame).
Solid line shows the best smooth broken power-law fit of $R$ points
(see text). {\em Bottom panel}: fractional residuals with respect
to the fit.}
\label{fig:all_lc}
\end{figure*}

Figure~\ref{fig:all_lc} shows the best-fit broken power laws, which are consistent
with no colour evolution.
The $R$ curve (the best sampled) cannot be fitted with
a single power law ($\chi^2/{\rm dof}=34/11$, significance of
$4\times10^{-4}$). We then tried to fit it with a smoothed broken power-law,
eq.~\ref{eq:bknpow} (Beuermann et al.~\cite{Beuermann99}) with the smoothness parameter
fixed to 1.
\begin{equation}
F(t)=\frac{F_0}{(t/t_{\rm b})^{\alpha_1}+(t/t_{\rm b})^{\alpha_2}}
\label{eq:bknpow}
\end{equation}
The best-fit parameters resulted: $\alpha_1=0.35\pm0.08$,
$\alpha_2=1.35\pm0.04$, $t_{\rm b}=11.5\pm0.5$~minutes
($\chi^2/{\rm dof}=10.0/9$; see solid line of Fig.~\ref{fig:all_lc}).
The value derived for the break time is consistent with that
found for the $R$ band by Butler~et~al.~(\cite{Butler06}).
The light curves of the other filters are not sampled comparably well
particularly at the beginning of observations.
We then fitted the other filters light curves by leaving only
the normalisation free to vary and fixing $\alpha_1$, $\alpha_2$,
$t_{\rm b}$ to their correspondent values found for the $R$ filter.
In all cases the $\chi^2$ turned out to be acceptable.
Figure~\ref{fig:all_lc} shows all these best-fit broken power laws,
therefore consistent with no colour evolution.
The bottom panel of Fig.~\ref{fig:all_lc}
shows the fractional residuals for each of the optical bands with respect
to their best-fit function, respectively.
The extrapolation of the $R$ light curve to $\sim1000$~minutes
after the burst is consistent within uncertainties with the observations
reported by Nanni~et~al.~(\cite{Nanni05}) in the $R_c$ filter,
ruling out the possibility of other breaks before $10^3$~minutes.
We note the possible presence of two humps in the
$B$ and $R$ light curves: the first comprises a 10--20\% excess above the
expected flux from the power law fit between 70 and 90 minutes and the
second occurs at 40--50 minutes with flux at 5--10\% in excess of the
fit.
If we repeat the fit by ignoring these points, the best-fit normalisation
decreases by $\sim$5\%, so that the corresponding excesses increase by
the same percentage.
Around 170~minutes we have a single point in $B$ which is $(25\pm10)$\%
in excess of the fit $B$ (2.5$\sigma$).
Interestingly, also Butler~et~al.~(\cite{Butler06}) found similar
residual variability in the optical bands at $\ge30$\% around the
best-fitting broken power-law model at $t>200$~minutes.
Alternatively, if we fix $\alpha_2=\alpha_{\rm x,\gamma}$, the fit
of the R light curve turns
out to be equally good ($\chi^2/{\rm dof}=10.3/10$), with
$\alpha_1=0.61\pm0.10$ and $t_{\rm b}=36\pm20$~minutes.
However, if we fit the light curves of the other filters by allowing just the
normalisation free to vary, we find acceptable fits,
except for $i'$, for which the fit is bad
($\chi^2/{\rm dof}=18.0/6$, chance probability of $6\times10^{-3}$).

\section{Spectral Energy Distribution}
\label{sec:SED}
\begin{figure*}
  \centering
\resizebox{\hsize}{!}{\includegraphics{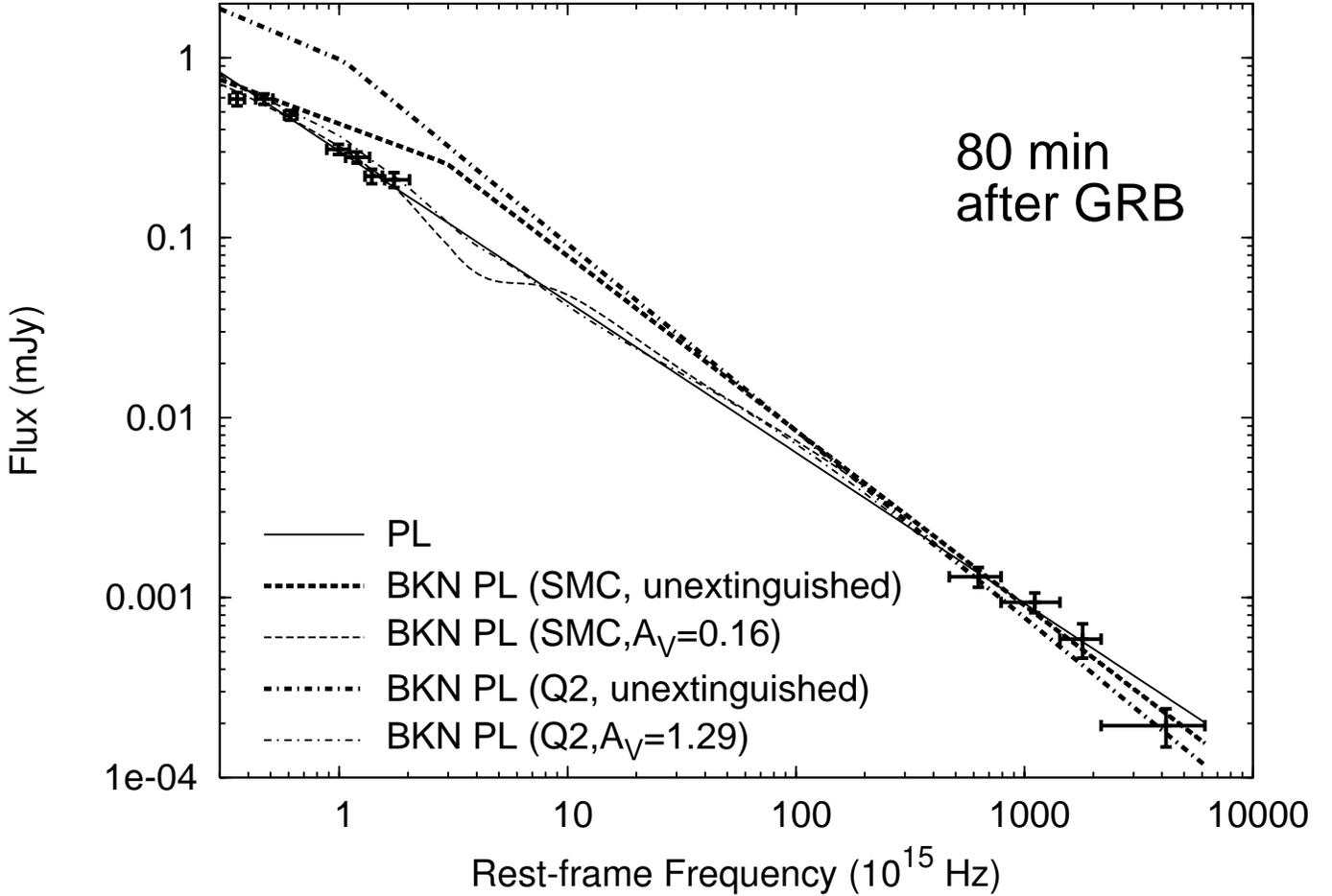}}
\caption{Rest-frame Spectral Energy Distribution at 80 min after the burst
from observed IR (Bloom et al.~\cite{Bloom05}) to X-rays measured with {\em Swift}/XRT.
A simple power law fits the spectrum with an index
$\beta_{\rm o,x}=0.84\pm0.02$
(thin solid line, $\chi^2/{\rm dof}=6.0/9$). Fluxes have been corrected
for Galactic extinction. Alternatively, the broken
power-law fit is shown when significant intrinsic extinction is
modelled: dashed line corresponds to the SMC profile, both extinguished
(thin line) and when the extinction is taken out (thick line).
Dashed-dotted line shows the result with a Q2 profile: extinguished
(thin line) and when the extinction is taken out (thick line).
See Table~\ref{tab:SED_fit}.}
\label{fig:SED}
\end{figure*}

Given the lack of evidence for dramatic colour changes during observations,
we found an epoch suitable to study the Spectral Energy Distribution (SED)
of the GRB afterglow. To this aim, we made use of IR observations reported
by Bloom et al.~(\cite{Bloom05}) in the $J$, $H$ and $K_s$ filters, carried out between
83.9 and 88.7 minutes after the GRB. The optimum time at which to calculate
the SED, when several measurements at different wavelengths were made
suitably closely in time, was chosen to be t$_{\rm SED}=80$~min.
We back-extrapolated the values reported by Bloom et al.~(\cite{Bloom05}) to $t_{\rm SED}$
assuming a power-law decay with $\alpha_2=1.35$ derived from fitting the
$R$ band (Sec.~\ref{sec:res_opt}). We corrected for a small Galactic
extinction, following Cardelli et al.~(\cite{Cardelli89}): $A_J=0.14$, $A_H=0.09$ and
$A_{Ks}=0.06$. We finally converted magnitudes into fluxes following
Campins et al.~(\cite{Campins85}). Optical $BVRi'$ magnitudes have been
interpolated linearly
to $t_{\rm SED}$ taking the measurements just before and after it.
X-ray observations began 15 minutes after $t_{\rm SED}$: we back-extrapolated
the X-ray flux assuming the simple best-fit power-law derived from X-ray
data, with $\alpha_{\rm x}=1.56\pm0.08$ (Sec.~\ref{sec:res_XRT}).
The errors on $\alpha_{\rm x}$ and normalisation, which are correlated,
have been propagated to estimate the uncertainty on the extrapolated X-ray flux.
The X-ray spectral shape is that of the spectrum
at $t=95$~min and rescaled according to the back-extrapolation factor.
We ignored rest-frame X-ray frequencies below $4.6\times10^{17}$~Hz
(i.e., 1.9~keV) because these are partially affected by significant $N_{\rm H}$.
The resulting SED is shown in Fig.~\ref{fig:SED}.

We fitted the whole SED with a simple power law, $F(\nu)\sim \nu^{-\beta}$:
$\beta_{\rm o,x}=0.84\pm0.02$ ($\chi^2/{\rm dof}=6.0/9$).
If we fit IR/optical and X-ray measurements with separate power laws, it
results: $\beta_{\rm o}=0.76\pm0.07$ and
$\beta_{\rm x}=\Gamma_{\rm x}-1=1.15\pm0.15$ ($\chi^2/{\rm dof}=4.3/7$).

We also fitted the SED with a broken power law with 
$\beta_{\rm o}-\beta_{\rm x} = 0.5$, assuming a number of extinction profiles
accounting for possible intrinsic extinction. The results are
discussed in Sec.~\ref{sec:SED_ext} and reported in Table~\ref{tab:SED_fit}.

\subsection{Element abundances and column densities}
\label{sec:met}
From the Keck optical absorption spectrum Penprase et al.~(\cite{Penprase06})
measure a Zn column density of $\log{(N_{\rm ZnII})}=13.58\pm0.15$~cm$^{-2}$.
Because of the redshift, the Lyman $\alpha$ feature does not fall within the
Keck optical band, so that the $N_{\rm HI}$ cannot be measured directly,
unlike other GRBs (see e.g. Vreeswijk et al.~\cite{Vreeswijk04}, Starling
et al.~\cite{Starling05}, Chen et al.~\cite{Chen05}, Fynbo et al.~\cite{Fynbo06}).
They assume a $\approx$1/10 solar metallicity typical of QSO-DLA systems,
i.e. [Zn/H]$\approx -1$ and find $N_{\rm H}\sim7.9\times10^{21}$~cm$^{-2}$.
The soft X-ray photoelectric cutoff measured in terms of $N_{\rm H}$ actually
is a measure of the metals column density $N_{\rm Z}$ under the assumption
of solar metallicity (Morrison \& McCammon~\cite{Morrison83}).
In particular, the soft X-ray
absorption is mainly due to $\alpha$ metals and secondly to the Fe-peak
metals, with no distinction between dust and gas. Oxygen K-shell absorption
edges represent the main contribution in the XRT energy band in the
assumption of relative solar abundances.
The relative contributions to the total absorption also depend on redshift,
given that the 0.2--10~keV energy band considered is fixed with respect
to the observer frame.
Hereafter we adopt the solar abundances as given by Asplund
et al.~(\cite{Asplund05}).
If we fit the X-ray spectrum by fixing the metallicity to be 1/10 solar,
the resulting $N_{\rm H}$ is significantly higher than that derived by
Penprase et al.~(\cite{Penprase06}) (see Table~\ref{tab:XRT_spec}).

If, however, we adopt a solar metallicity, the $N_{\rm H}$
derived from X-rays, $8.0_{-2.6}^{+3.3}\times10^{21}$~cm$^{-2}$, is 
consistent with the value derived by Penprase et al.~(\cite{Penprase06}),
although based on an incompatible metallicity assumption.
Possibly, this could be explained by a significant fraction of molecular
gas: in fact, Arabadjis \& Bregman~(\cite{Arabadjis99})
and Baumgartner \& Mushotzky~(\cite{Baumgartner06}) found that
in the Milky Way for column densities higher than $10^{21}$~cm$^{-2}$
the X-ray column densities are 1.5--3 times as high as those derived
from radio/optical. 
Alternatively, Zn depletion into dust could possibly explain this. However, apart
from the cool dense ISM, in which the zinc fraction in dust grains
can be as high as 20\% ISM (Savage \& Sembach~\cite{Savage96}), Zn depletion is
usually negligible and no evidence for it has been found in GRB-DLAs
so far. The cool dense ISM is incompatible with the depletion pattern
of the warm disk found by Penprase et al.~(\cite{Penprase06}; see below, this Section).
Another possibility to reconcile the mismatch between metallicity and
X-ray absorption is to allow X-ray absorbing elements to
be overabundant with respect to the measured Zn: e.g,
an overabundance of $\alpha$ metals with respect to zinc,
as it appears to be the case for GRB~050401 (Watson et al.~\cite{Watson06}).
However, in the case of GRB~051111 it would be difficult to explain why
silicon is not very abundant: from table~2 of Penprase et al.~(\cite{Penprase06})
we take $\log{({\rm Si})}=16.18\pm0.20$ and $\log{({\rm Zn})}=13.58\pm0.15$.
It follows that [Si/Zn]$=\log{({\rm Si}/{\rm Zn})}-\log{({\rm Si}/{\rm Zn})_{\odot}}=-0.30\pm0.25$
and this could hardly be accounted for as due to dust depletion only.
Another possibility is an overabundance of oxygen, as the $\alpha$
element mainly responsible for the X-ray absorption.
We measure the O abundance directly and we compare it with that of Zn,
which we know from the Keck optical spectrum because it is non-refractory
to dust depletion.
From the $N_{\rm H}=8.0_{-2.6}^{+3.3}\times10^{21}$~cm$^{-2}$
measured assuming solar metallicity, we derive an O column density of
$\log{({\rm O})}=18.56\pm0.18$, i.e.
[O/Zn]$=\log{({\rm O}/{\rm Zn})}-\log{({\rm O}/{\rm Zn})_{\odot}}=0.9\pm0.2$.
The O column density value estimated assuming
$N_{\rm H}=7.1_{-2.4}^{+3.2}\times10^{22}$~cm$^{-2}$ (from Table~\ref{tab:XRT_spec})
and a metallicity of 1/10 solar is equivalent within the errors:
the oxygen is 10 times less abundant and the $N_{\rm H}$ column density
is $9\pm5$ times higher.
If we alternatively assume the lowest value we obtained, i.e. in the 0.3--10~keV
ignoring the channel from 0.43 to 0.6~keV, 
$N_{\rm H}=5.1_{-2.1}^{+2.9}\times10^{21}$~cm$^{-2}$, the relative oxygen
abundance with respect to zinc turns out to be [O/Zn]=$0.7\pm0.3$.

The value of [O/Zn] can be compared with [$\alpha$/Zn]$=0.5\pm0.2$ found by
Watson et al.~(\cite{Watson06}) for GRB~050401.
If this is also true for GRB~051111, we do not know a mechanism that
can explain the observed overabundance of oxygen with respect to other
$\alpha$ elements, e.g. either [O/Si]=$1.2\pm0.3$ ([O/Si]=$1.0\pm0.4$)
assuming $N_{\rm H}=8.0_{-2.6}^{+3.3}\times10^{21}$~cm$^{-2}$
($N_{\rm H}=5.1_{-2.1}^{+2.9}\times10^{21}$~cm$^{-2}$).

As we do not have a direct measure of either O or H,
we have to assume both [O/H] and [Zn/H], provided that
eq.~\ref{eq:abund} is satisfied:
\begin{equation}
[{\rm O}/{\rm H}] - [{\rm Zn}/{\rm H}] = [{\rm O}/{\rm Zn}]
\label{eq:abund}
\end{equation}
From the low column densities of Fe, Cr, Si and Mn compared with
Zn, Penprase et al.~(\cite{Penprase06}) infer a strong dust depletion, in agreement
with what found for other GRBs (Savaglio \& Fall~\cite{Savaglio04}).
The depletion pattern matches that of a warm disk (WD)
and marginally that of a warm disk plus halo (WDH) observed in the Milky Way
(see e.g. Savaglio et al~\cite{Savaglio03} and references therein).
From the 1/10 solar metallicity assumption, Penprase et al.~(\cite{Penprase06})
estimate a dust-to-gas ratio of about 1/12 that of the Milky Way and from their value
of the $N_{\rm H}$ they derived an extinction of $A_V\approx0.55$.

It must be pointed out that the measure of the ZnII column density by
Penprase et al.~(\cite{Penprase06}) has not universally been accepted
(Prochaska~\cite{Prochaska05,Prochaska06a}, Prochaska et al.~\cite{Prochaska06b});
in particular, according to Prochaska et al.~(\cite{Prochaska06b}) both ZnII and
SiII are saturated and they derive a lower limit on the amount of zinc:
$\log{(N_{\rm ZnII})}>13.71$~cm$^{-2}$,
i.e. about 1$\sigma$ above what obtained by Penprase et al.~(\cite{Penprase06}).
Here, we do not discuss the reasons for these different conclusions;
rather, we note that in this case it would be [O/Zn]$<0.8$ and
the evidence for overabundant oxygen would be much less compelling.
Also the optical extinction mentioned above would
become merely a lower limit: $A_V>0.55$.

\subsection{Host extinction}
\label{sec:SED_ext}
According to the afterglow broad band spectrum expected in the
fireball model (Sari et al.~\cite{Sari98}), we assume a straight power law,
$F_{\nu}\sim \nu^{-\beta}$, or a broken power law with
$\Delta\beta=0.5$ in the typical scenario of slow cooling with
the cooling frequency $\nu_{\rm c}$ lying between optical and X-rays
(see Sec.~\ref{sec:disc}).
However, the possible dust extinction due either to the circumburst
environment and/or the host galaxy not only suppresses the optical
flux but also reddens it in a way which reflects
the properties of dust. As a result, $\beta_{\rm o}$
measured in the optical domain may be significantly steeper than
the intrinsic one, $\beta_{\rm o,i}$,
we would have measured in the absence of extinction ($A_V=0$).
The values for $A_V$ and $\beta_{\rm o,i}$ as estimated from fitting
the SEDs of GRBs are to some degree correlated, as proved by previous
studies: e.g., see Panaitescu~(\cite{Panaitescu05}),
Stratta et al.~(\cite{Stratta04}) and Kann et al.~(\cite{Kann06}).
In order to account for possible intrinsic extinction, we fitted the
SED with the following general law (eq.~\ref{eq:ext}):
\begin{equation}
F_\nu = F_0(\nu)\ 10^{-0.4 A(\nu)}
\label{eq:ext}
\end{equation}
where $F_0(\nu)$ is the afterglow spectrum in the absence of extinction:
$F_0(\nu)\propto\nu^{-\beta_{\rm o,i}}$ ($\beta_{\rm x}=\beta_{\rm o,i}$)
in the power-law case, $F_0(\nu)\propto\nu^{-\beta_{\rm o,i}}$
(for $\nu<\nu_0$) and $F_0(\nu)\propto\nu^{-\beta_{\rm x}}$
($\beta_{\rm x}=\beta_{\rm o,i}+0.5$, for $\nu>\nu_0$) in the
broken power-law case ($\nu_0$ is the break frequency).
$A(\nu)$ is the extinction profile as a function of $\nu$ and
depends on the dust model assumed.

We tried to account for intrinsic extinction assuming nine different
extinction profiles:
\begin{enumerate}
\item Milky Way (MW) using the parameterisation by Pei~(\cite{Pei92}),
yielding $N_{\rm H}/A_V=1.6\times10^{21}$~cm$^{-2}$.
\item Large Magellanic Cloud (LMC) using the parameterisation by
Pei~(\cite{Pei92}), yielding $N_{\rm H}/A_V=7.6\times10^{21}$~cm$^{-2}$.
\item Small Magellanic Cloud (SMC) using the parameterisation by
Pei~(\cite{Pei92}), yielding $N_{\rm H}/A_V=1.5\times10^{22}$~cm$^{-2}$.
\item The ``Q1'' curve by Maiolino et al.~(\cite{Maiolino01}) derived for
population of dust grains whose size distribution is given by a power law,
$dn/da\propto a^{-q}, (a_{\rm min}\le a\le a_{\rm max})$, where $q=3.5$,
$a_{\rm min}=0.005$~$\mu$m, $a_{\rm max}=10$~$\mu$m. It is
$N_{\rm H}/A_V=6.7\times10^{21}$~cm$^{-2}$ assuming a standard
gas-to-dust ratio in the ISM.
\item The ``Q2'' curve by Maiolino et al.~(\cite{Maiolino01}) derived
under the same assumptions as for the Q1 profile, but the following parameters:
$q=2.5$ $a_{\rm max}=1$~$\mu$m. It is
$N_{\rm H}/A_V=3.3\times10^{21}$~cm$^{-2}$ assuming a standard
gas-to-dust ratio in the ISM.
\item SN extinction profile taken from Maiolino et al.~(\cite{Maiolino04})
for the case of a $10 M_{\odot}$ SN with solar metallicity (for our purposes this
case is similar to the other SN profiles discussed by those authors).
\item The type I extinction law (hereafter CLW1) provided by
Chen et al.~(\cite{Chen06}), derived assuming a dust mainly composed by graphite
and silicate grains following a power law multiplied by an exponential,
$dn/da\propto a^{-q} e^{a/a_c}, (a_{\rm min}\le a\le a_{\rm max})$,
where $a_{\rm min}=0.05$~$\mu$m, $a_{\rm max}=2.5$~$\mu$m,
$q=2.61$ and $a_{c}=0.21$~$\mu$m.
\item The type II extinction law (hereafter CLW2) provided by
Chen et al.~(\cite{Chen06}), under the same assumptions as for CLW1, but the
following parameters: $q=3.09$ and $a_{c}=0.29$~$\mu$m. 
\item This extinction law corresponds to a power law,
$A(\nu)=A_V (\nu/\nu_V)^\gamma$, as proposed
by Savaglio \& Fall~(\cite{Savaglio04}) for GRB~020813. Since
the fitting procedure in the case of our data does not constrain
very well the two parameters $A_V$ and $\gamma$, we fixed $\gamma$
in two boundary cases found by Savaglio \& Fall~(\cite{Savaglio04}),
i.e. $\gamma=0$ and $\gamma=0.85$. Hereafter the two cases are referred to as
SF($\gamma=0$) and SF($\gamma=0.85$).
\end{enumerate}
\begin{figure}
  \centering
\resizebox{\hsize}{!}{\includegraphics{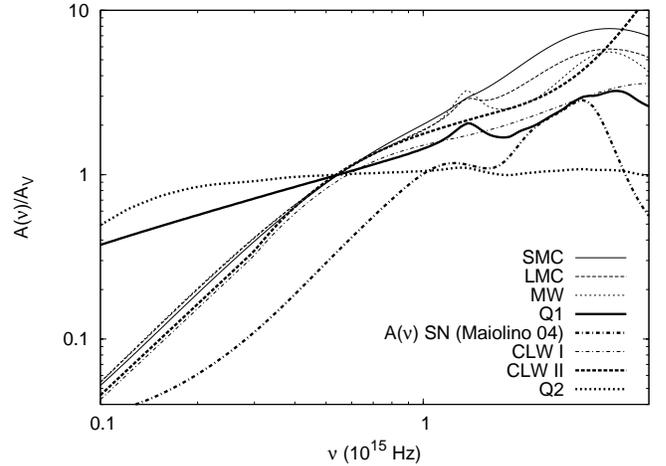}}
\caption{Extinction profiles of eight different laws: SMC (thin solid line),
LMC (thin dashed line), MW (thin dotted line), Q1 (thick solid line),
the absolute extinction expected for a $10 M_{\odot}$ solar metallicity
SN profile from Maiolino et al.~(\cite{Maiolino04}) (thick dashed-dotted line),
CLW1 (thin dashed-dotted line), CLW2 (thick dashed line),
Q2 (thick dotted line).}
\label{fig:extprof}
\end{figure}

The first eight extinction curves listed above considered are shown in
Fig.~\ref{fig:extprof}. The results are reported in Table~\ref{tab:SED_fit}.
\begin{table*}
\caption{Best-fit results of the IR-optical-X SED at
  $t=80$~minutes. Uncertainties are at 90\% CL.}
\label{tab:SED_fit}
\centering
\begin{tabular}{llcccr}
\hline\hline
Model$^{(a)}$ & Extinction & $A_V^{(b)}$ & $\beta_{\rm o,i}$ & $\nu_0^{(c)}$ &
$\chi^2/{\rm dof}$\\
 & Law &   &  & ($10^{15}$~Hz) &\\
\hline
PL     & Unextinguished &     0   & $0.84\pm0.02$          & --       & 6.0/9\\
PL     & SMC            & $<0.02$ & $0.84\pm0.02$          & --       & 13.6/8\\
PL     & LMC            & $<0.02$ & $0.84\pm0.02$          & --       & 13.6/8\\
PL     & MW             & $<0.03$ & $0.84\pm0.02$          & --       & 13.6/8\\
PL     & Q1             & $<0.05$ & $0.84\pm0.02$          & --       & 13.6/8\\
PL     & Q2             & $<0.19$ & $0.84\pm0.03$          & --       & 13.6/8\\
PL     & CLW I          & $<0.03$ & $0.84\pm0.02$          & --       & 13.6/8\\
PL     & CLW II         & $<0.03$ & $0.84\pm0.02$          & --       & 13.6/8\\
PL     & SN 10 $M_{\odot}$ & -- & $0.88\pm0.02$          & --       & 225/9\\
PL     & SF($\gamma=0.0$)  & $<0.34$ & $0.84\pm0.04$          & --       & 13.6/8\\
PL     & SF($\gamma=0.85$) & $<0.03$ & $0.84\pm0.02$          & --       & 13.6/8\\
BKN PL & Unextinguished &     0   & $0.70_{-0.01}^{+0.09}$ & $134_{-94}^{+312}$ & 12.95/8\\
BKN PL & SMC            & $0.16_{-0.11}^{+0.06}$ & $0.47_{-0.05}^{+0.18}$ & $3.0_{-1.6}^{+52}$ & 7.7/7\\
BKN PL & LMC            & $0.21_{-0.13}^{+0.05}$ & $0.44_{-0.03}^{+0.18}$ & $1.8_{-0.6}^{+30}$ & 6.4/7\\
BKN PL & MW             & $0.20_{-0.05}^{+0.06}$ & $0.43_{-0.03}^{+0.20}$ & $1.5_{-0.5}^{+30}$ & 6.4/7\\
BKN PL & Q1             & $0.39_{-0.10}^{+0.11}$ & $0.46_{-0.05}^{+0.18}$ & $1.5_{-0.5}^{+34}$ & 6.7/7\\
BKN PL & Q2             & $1.29_{-0.45}^{+0.53}$ & $0.54_{-0.06}^{+0.07}$ & $1.1_{-0.1}^{+1000}$ & 8.8/7\\
BKN PL & CLW I          & $0.30_{-0.28}^{+0.09}$ & $0.48_{-0.09}^{+0.22}$ & $2.9_{-1.9}^{+1000}$ & 9.8/7\\
BKN PL & CLW II         & $0.25_{-0.20}^{+0.08}$ & $0.47_{-0.08}^{+0.20}$ & $2.4_{-1.4}^{+1000}$ & 8.8/7\\
BKN PL & SN 10 $M_{\odot}$ & -- & $0.46\pm0.02$ & $1.7_{-0.4}^{+0.3}$ & 78/8\\
BKN PL & SF($\gamma=0.0$)  & $1.8_{-0.6}^{+0.9}$ & $0.60_{-0.08}^{+0.12}$ & $1.0_{-0.0}^{+1000}$ & 11.4/7\\
BKN PL & SF($\gamma=0.85$) & $0.26_{-0.18}^{+0.08}$ & $0.46_{-0.03}^{+0.18}$ & $2.0_{-0.7}^{+42}$ & 7.5/7\\
\hline
\end{tabular}\\
\flushleft
$^{(a)}$ For the power-law (PL) model, we assumed $\beta_{\rm x}=\beta_{\rm o,i}$.
For the broken power-law (BKN PL) model, we assumed $\beta_{\rm x}-\beta_{\rm o,i}=0.5$.\\
$^{(b)}$ $A_V$ was left free to vary in the [0-10] interval.\\
$^{(c)}$ $\nu_0$ is the break frequency for the BKN PL model.
It was left free to vary in the $(1-1000)\times10^{15}$~Hz interval.
\end{table*}
%

\section{Discussion}
\label{sec:disc}
In order to interpret the light curves and spectra at different
epochs for different energy bands according to the standard fireball model,
we consider the two usual cases of different medium density profile,
$\rho\propto r^{-k}$: ISM (Sari et al.~\cite{Sari98}) ($k=0$) vs.
Wind (Chevalier \& Li~\cite{Chevalier99}) ($k=2$).
Hereafter, $\nu_m$ and $\nu_c$ are
the peak synchrotron energy corresponding to the minimum energy of
the electron energy distribution ($dN/d\gamma\propto \gamma^{-p}$,
for $\gamma>\gamma_m$, where $p$ is the electron energy distribution
index) and the cooling frequency, where the electrons lose their energy
within the dynamical timescale, respectively.
Likewise, $\nu_{\rm o}$ and $\nu_{\rm x}$ are the optical and X-ray
frequencies, respectively.

First of all, we briefly discuss whether a break frequency lies
between optical and X-rays at 80 minutes. As reported in
Sec.~\ref{sec:SED}, a single power-law fit from optical to X-rays
is acceptable with $\beta_{\rm o,x}=0.84\pm0.02$.
However, despite the goodness of the fit, the
optical and X-ray indices alone, $\beta_{\rm o}=0.76\pm0.07$
and $\beta_{\rm x}=1.15\pm0.15$, seem to support a break consistent
with $\beta_{\rm x}-\beta_{\rm o}=0.5$, often expected in the
fireball model. Hence, although the single power law cannot
be ruled out at all, it seems disfavoured.
Notably, in the case of a single power law, an important
consequence would be the negligible optical extinction: $A_V<0.05$
(90\% CL) for almost all of the extinction profiles considered (see
Table~\ref{tab:SED_fit}), unless one assumes very flat profiles
around $1\times10^{15}$~Hz, such as the Q2 and SF($\gamma=0$)
(see Fig.~\ref{fig:extprof}) for which it is $A_V<0.19$ and
$A_V<0.34$, respectively.
Given the high column densities of metals and the significant
X-ray absorption measured, this appears to be hardly the case.
In contrast, a broken power law seems to better allow for a higher
extinction: the models based on big dust grains give
values of $A_V$ between 0.25 and 1.29 and, in particular, the
grey extinction (Stratta et al.~\cite{Stratta04}) power-law profiles considered
by Savaglio \& Fall~(\cite{Savaglio04}) can account for even higher values,
ranging from 0.26 to 1.8.

Hence the broken power-law case seems to be more consistent with the
amount of extinction expected from the significant dust depletion measured
in the optical spectrum: $A_V\sim0.55$ or even much higher,
as discussed in Sec~\ref{sec:met}.  

From these considerations, in the following we consider the
alternative of a break frequency between optical and X-rays,
which appears to be more likely for the SED at 80~minutes.

\begin{itemize}
\item {\em $\nu_{\rm x}<\nu_c$}

At $t_{\rm SED}$, i.e. after the break in the optical light curve,
the case $\nu_m<\nu_{\rm x}<\nu_c$ is unlikely to describe the data.
If $\nu_{\rm o}<\nu_m<\nu_{\rm x}<\nu_c$, we should
expect $\beta_{\rm o,i}=-1/3$ (all cases), which is inconsistent
with our results (Table~\ref{tab:SED_fit}).
Alternatively, if $\nu_m<\nu_{\rm o}<\nu_{\rm x}<\nu_c$,
then it must be $\beta_{\rm o,i}=\beta_{\rm x}$. From
Table~\ref{tab:SED_fit} the only solution is the case of a single
power law from optical to X-ray and it is $\beta_{\rm o,i}=\beta_{\rm x}=0.84\pm0.02$.
For both $k=0,2$ cases, it should be $\beta_{\rm x}=(p-1)/2$,
implying $p=2.68\pm0.04$. The measured value for $\alpha_{\rm x}$ of $1.56\pm0.08$
is inconsistent with that expected for the ISM case, namely $3(p-1)/4=1.26\pm0.03$,
but is very  marginally consistent (2\% significance) with the Wind case:
$(3p-1)/4=1.76\pm0.03$.
This, combined with the negligible extinction implied by the single
power law case (Table~\ref{tab:SED_fit}), makes this scenario
unlikely.

\item {\em $\nu_c<\nu_{\rm x}<\nu_m$} (fast cooling)

In all cases, it should be $\beta_{\rm x}=1/2$, clearly incompatible
with data.

\item {\em $\nu_c<\nu_m<\nu_{\rm x}$} (fast cooling)

In this case, following Sari et al.~(\cite{Sari98}) we should expect a previous
break in the X-ray light curve from $\alpha_{\rm x}=1/4$ to
$\alpha_{\rm x}=(3p-2)/4$ when $\nu_m$ crossed the X-ray band.
Furthermore, when at earlier epoch $\nu_c$ crossed the X-ray band,
the X-ray light curve should have been even shallower, with
$\alpha=-1/6$. All this is at odds with the X-ray light
curve inferred from Fig.~\ref{fig:all_lc_x}.

\item {\em $\nu_m<\nu_c<\nu_{\rm x}$} (slow cooling)

In both cases, it is $\beta_{\rm x}=p/2$, yielding
$p=2.3\pm0.3$. This implies $\alpha_{\rm x}=(3p-2)/4=1.22\pm0.22$.
This is compatible with the measured $1.56\pm0.08$ within 1.5 $\sigma$,
although somewhat shallower.
The case of $\nu_{\rm o}<\nu_m$ has already been ruled out because the
expected optical decay ($\alpha=-1/2$ or $0$ for $k=0,2$, respectively)
is too shallow compared with the measured decay ($\alpha_2$).
Hence, hereafter we consider the case $\nu_m<\nu_{\rm o}<\nu_c<\nu_{\rm x}$.
For both cases, it is $\beta_{\rm o,i}=(p-1)/2=0.65\pm0.15$. This 
value is in agreement with most of those obtained by fitting the SED
(see Table~\ref{tab:SED_fit}).
The expected optical decay power-law index is
$\alpha_{\rm o}=3(p-1)/4=1.0\pm0.2$ (ISM) and 
$\alpha_{\rm o}=(3p-1)/4=1.5\pm0.2$ (Wind), which favours the Wind profile,
given that $\alpha_2=1.35\pm0.04$.
In the Wind case, the optical decay should
be steeper than the X-ray decay by 0.25; the observed difference,
$\alpha_2-\alpha_{\rm x}=-0.2\pm0.1$, could be indicative of a stratified
wind, but the observed pre-break optical decay would be difficult to
explain in this scenario. See Sec.~\ref{s:disc_singlex}.
\end{itemize}

\subsection{Break in the optical light curve}
The interpretation of the optical break as the result of the passage of the
cooling frequency across the optical bands appears to be unlikely,
because we observe $\Delta\alpha\sim1$ to be compared with 0.25
expected in that case. The passage of $\nu_m$ cannot explain it
either, as the expected pre-break slope should be positive 
(specifically, $\alpha_1=-0.5$), which is definitely not the case.
The break is unlikely due to a jet, given the different behaviour
exhibited by the X-ray afterglow decay, which seems to rule out
the achromatic nature, required by the jet scenario, not to mention
that a break time of $\sim12$~minutes would be unusually early.
In the following we discuss the optical break in comparison with
the X-ray behaviour in an attempt to explain it in a single framework.

\subsubsection{Breaks in the X-ray power-law decay}
If we consider the connection between the $\gamma$-ray
tail of the prompt emission and the late X-ray afterglow as due to chance,
it is natural to assume that X-rays mimic the optical decay: in this
case, after the prompt emission the X-ray decay would be steeper than
$\alpha_{{\rm x},\gamma}=1.54$, followed by a shallow decay
phase and finally reconnecting to the normal decay measured 
$\alpha_{\rm x}=1.56$.
This seems consistent with the slope observed during the tail of
the prompt emission, $\alpha_{\gamma}=1.8\pm0.2$, which is
marginally steeper than $\alpha_{{\rm x},\gamma}$.
The steep-shallow-normal behaviour has been observed with {\em Swift}
for many early X-ray afterglows. The steep decay in the early
phase is interpreted as tail of the prompt emission and not produced
by the interaction of the fireball with the surrounding medium,
as for the afterglow (Kumar \& Panaitescu~\cite{Kumar00}; Tagliaferri
et al.~\cite{Tagliaferri05}; Lazzati \& Begelman~\cite{Lazzati06}).
The following shallow X-ray decay would be the result of a continuous
energy injection at the end of which there is a steepening
in the light curve (Nousek et al.~\cite{Nousek06}; Zhang et al.~\cite{Zhang06}).
In this case, GRB~051111 would represent one of the few cases
in which this behaviour, typical of early X-ray afterglows,
is seen in the early optical afterglow.
This scenario for GRB~051111 seems to be supported also by the
ROTSE observations (Yost et al.~\cite{Yost06}) as well as
by the data taken with the Katz Automatic Imaging Telescope (KAIT)
reported by Butler~et~al.~(\cite{Butler06}).

The alternative description of the X-ray light curve in terms of a double
broken power model reported in Sec.~\ref{sec:res_XRT} fits the possible
X-ray bump visible around 600~minutes, followed by a possible steepening.
In principle, these variations with respect to the simple power-law fit may
be explained in terms of a minor late energy injection process, responsible
for a short-lived shallow decay, or alternatively, as due to the narrow
component of a two-component jet (e.g. Huang et al.~\cite{Huang04},
Granot~\cite{Granot05}). However in the latter case from simultaneous and
late optical observations (Butler et al.~\cite{Butler06},
Nanni et al.~\cite{Nanni05}) the expected steepening after the putative
second break at $\sim600$ minutes is not evident. Unfortunately, the late
X-ray upper limit shown in Fig.~\ref{fig:all_lc_x} is not conclusive.

\subsubsection{Single X-ray power-law decay}
\label{s:disc_singlex}
Alternatively, as shown in Sec.~\ref{sec:res_BATXRT}, one may consider
that the tail of the prompt event
connects with the late X-ray afterglow light curve with a single power-law
slope of $\alpha_{{\rm x},\gamma}=1.54\pm0.10$.
Although the X-ray data gap until XRT observations do not exclude breaks
in the power law as found for many {\em Swift} early
X-ray afterglows (O'Brien et al.~\cite{OBrien06}), it seems unlikely that in case of
variations the two data sets would
connect so well despite 2--3 orders of magnitude difference in fluxes.
A single power law connecting the BAT $\gamma$-ray tail with the XRT
afterglow has already been observed for some GRBs: e.g. GRB~050721, 
GRB~050726, GRB~050801 and GRB~050802, see fig.~3 of O'Brien et al.~(\cite{OBrien06}).
Also for the very first GRB for which an afterglow was found,
GRB~970228, the X-ray power-law back-extrapolation to the time of the
burst nicely connects with the X-ray prompt emission (Costa et al.~\cite{Costa97}).
Therefore, in this section we discuss the case in which there is no break in
the X-ray light curve simultaneous to that in the optical band.
The different decay rates of optical and X-rays require the presence
of a break frequency in between, as already argued above.
Here we also assume that there is no significant spectral X-ray evolution
from the prompt emission tail to the XRT, i.e. $\beta_{\rm x}$ being
constant. This is not at odds with $\beta_{\gamma}=0.44$ measured
during the $\gamma$-ray tail: e.g., in the case of GRB~050726 we have
$\beta_{\gamma}=0.01\pm0.17$ and $\beta_{\rm x}=0.94\pm0.07$
(O'Brien et al.~\cite{OBrien06}).

A possible explanation of the optical break with no corresponding break
in the X-ray light curve is when the afterglow front shock passes from
ISM to a Wind environment.
In this case we would expect a steepening of 0.5 in the optical bands only
(because $\nu_c<\nu_{\rm x}$). However the measured $\Delta\alpha$ is
about 1. If we consider the more general case of density scaling
as $r^{-k}$, the expected steepening due to passage from ISM to
Wind is $\Delta\alpha=k/(8-2k)$, i.e. $k\approx2.7$.
This solution has been suggested to explain the similar case of GRB~060206
in which an optical break with no corresponding X-ray break was observed
(Monfardini et al.~\cite{Monfardini06}).
However, in this case ($k>2$) the X-ray decay $\alpha_{{\rm x},\gamma}$
should be smaller by at least $0.25$ than the optical $\alpha_2$, while we find
$\alpha_{{\rm x},\gamma}-\alpha_2=0.19\pm0.11$.

Alternatively, the post-break decay index (optical and X-ray) themselves
can be explained if we assume $p=2.75$ and $k=1/2$. The essential problem for the
ISM-wind transition (stratified wind) model is that the pre-break optical
decay index $\alpha_1=0.35$ is too shallow. 
In a more generalised model, a stratified wind model might include a 
transition from $k<0$ (increasing density with radius) to $k>0$.
Compared with the ISM ($k=0$), this should give a shallower pre-break optical decay. 
From table~5 in Yost et~al. (\cite{Yost03}), in the case $\nu_m<\nu_{\rm o} < \nu_{\rm c}$,
the decay index $\alpha_{\rm o}$ for the Wind model is given by 
$\alpha_{\rm o}=3p/4-[(12-5k)/(16-4k)] > 3p/4-5/4 = 0.81$ ($p=2.75$)
or $0.48$ ($p=2.3$).
Therefore the observed pre-break optical decay is too shallow to satisfy
this scenario.

A transition from ISM to a wind environment in principle
may occur when, for instance, a He merger progenitor, beginning
with a massive star with strong winds, travels up to 10~pc
from the wind environment before the GRB explosion
(Fryer et al.~\cite{Fryer06}). In this scenario, the progenitor
should have run away from the line of sight to the observer.
Alternatively, the possibility
of a binary system with a Wolf-Rayet star seems more unlikely
to produce such an environment, because of the strong wind
dominating that of the companion (van~Marle et al.~\cite{vanMarle06}),
unless one considers a double Wolf-Rayet system, which would be very rare.

Finally, in this scenario, there is another aspect hard to reconcile
with the overall picture: if the tail of the prompt emission is
the beginning of the afterglow, then the $\gamma$-ray spectral
index, $\beta_{\gamma}=0.44$, and the power-law decay index,
$\alpha_{{\rm x},\gamma}=1.54$, appear to be incompatible with the
synchrotron shock model interpreted as produced by external shocks.

\subsection{The $\gamma$-ray tail}
Interestingly, Giblin et al.~(\cite{Giblin02}) studied the $\gamma$-ray tail
of a number of FRED-like GRBs detected with BATSE (Paciesas et al.~\cite{Paciesas99}).
They tested whether the decay
rate and the spectrum are consistent with the expectations of an early
high-energy afterglow synchrotron emission due to external shocks.
Their results show that about 20\% of the GRBs considered are consistent
with it, in particular with the evolution expected for a jet.

In the jet case, the power-law decay index, $\alpha_{\gamma}=1.8$,
should correspond to $p$, provided that $p>2$. As it is not the case,
we have to apply the formalism by Dai \& Cheng~(\cite{Dai01}) for $p<2$,
according to which we have two possible closure relations:
$4\alpha_{\gamma}-2\beta_{\gamma} -7=0$ ($\nu_\gamma<\nu_c$), or
$4\alpha_{\gamma}-2\beta_{\gamma} -6=0$ ($\nu_\gamma>\nu_c$).
In either case, they turn out to be acceptable within uncertainties:
$-0.7\pm0.8$ yielding $p=1.9\pm0.1$ and $0.3\pm0.8$ yielding $p=0.9\pm0.1$,
for the $\nu_\gamma<\nu_c$ and $\nu_\gamma>\nu_c$ cases, respectively.
This result shows that the tail of the prompt emission is
consistent with the jet case, as found by Giblin et al.~(\cite{Giblin02})
for some FREDs by BATSE. It must be pointed out, however, that the values
for $p$ in this case seem incompatible with $2.3\pm0.3$
derived from the optical/X-ray afterglow.

Curvature effects alone cannot account for the
$\alpha_\gamma$ and $\beta_\gamma$ measured: in this case the
expected closure relation $\alpha_\gamma-\beta_\gamma -2=0$
(e.g., Dermer~\cite{Dermer04})
turns out to be $-0.6\pm0.2$, incompatible with zero.

Notably, one of the GRBs studied by Giblin et al.~(\cite{Giblin02}),
GRB~910602, is strikingly similar to GRB~051111 in both temporal decay and spectrum:
$\alpha_{\gamma}=1.74$ and $\beta_{\gamma}=0.52$. According to the
classification of Giblin et al.~(\cite{Giblin02}), this is a FRED with multi pulses
near the peak. Also for this GRB, there is no softening along the
tail and it appears to be inconsistent with the evolution of a
spherical blast wave.

\section{Conclusions}
\label{sec:conc}
We provided a detailed study of the properties of both
prompt emission and multi-wavelength early afterglow of a
typical FRED burst detected by {\em Swift}.
The optical afterglow light curve is fitted by a smooth broken power law,
while little can be conclusively said about the simultaneous
X-ray light curve. Interestingly, the flux of the $\gamma$-ray
tail of the prompt emission, extrapolated in the 0.2--10~keV
energy band, matches the back-extrapolation of the late
X-ray afterglow with $\alpha_{{\rm x},\gamma}=1.54\pm0.10$.
We discussed two alternative cases, based on the behaviour
of the X-ray afterglow light curve soon after the prompt
event until $\sim$100 minutes, when XRT began observing. 

If the connection between the tail of the
prompt event and the late X-ray afterglow is pure chance,
the $\gamma$-ray tail is the result of internal shocks
and has no connection with the late X-ray
afterglow. The X-ray afterglow light curve could mimic 
that measured in the optical, which monitors the shallow-to-normal
decay phase observed in several early X-ray afterglows with
{\em Swift} and interpreted as the gradual end of a continuous
energy injection process into the front shock of the afterglow.
In this case, this would be one of the first bursts, whose
early optical afterglow shows the typical behaviour seen
in X-rays. This interpretation seems to be consistent with
other data sets (Butler et~al. \cite{Butler06};
Yost et~al. \cite{Yost06}).

Alternatively, if the matching between $\gamma$-ray tail and
the late X-ray afterglow is evidence for an early beginning of the X-ray
afterglow during the tail of the prompt event, it is
therefore sensible to assume no breaks in the X-ray light curve,
while in contrast, the optical exhibits a break
around 12 minutes after the burst. The break in the light
curve observed only in the optical with $\Delta\alpha\sim1$
could be the result of
a passage from ISM to a Wind environment, $\rho\propto r^{-k}$,
with $k\approx2.7$.
A weak point of this scenario is the fact that after the break
the X-ray decay should be shallower than that of the optical
by $0.25$ whereas we observe the optical
decay index to be marginally {\em steeper} by $0.19\pm0.11$.
Alternatively, a stratified wind with $p=2.75$ and passing from
a region of $k<0$ to $k>0$ could account for this, but
the observed pre-break optical decay appears unaccountably shallow.

From the SED at 80 minutes, we find that the presence of the
cooling frequency between optical and X-ray is consistent
with significant intrinsic dust extinction inferred from
the optical spectrum (Prochaska~\cite{Prochaska05}; Penprase
et al.~\cite{Penprase06}):
$0.16\lesssim A_V\lesssim 0.21$ for standard extinction profiles
(MW, SMC, LMC), or more probably $0.25\lesssim A_V\lesssim 1.8$
for dust dominated by big grains
(Maiolino et al.~\cite{Maiolino01}; Chen et al.~\cite{Chen06})
or grey extinction
laws, as found for other GRBs (Stratta et al.~\cite{Stratta04};
Savaglio \& Fall~\cite{Savaglio04}).
These kinds of extinction profiles might also be connected with
dust destruction within a few tens of
parsec of the GRB progenitor due to the intense UV and X-rays
of the prompt emission, as suggested by several 
authors (Waxman \& Draine~\cite{Waxman00}; Fruchter et al.~\cite{Fruchter01};
Perna et al.~\cite{Perna03}).

In contrast, the absence of a spectral break between optical
and X-ray would constrain the dust content ($A_V<0.19$),
thus making it hard to reconcile with the column densities
measured in the optical spectrum. Furthermore, it would
also be hard to explain the different decay in X-rays
($\alpha_{\rm x}=1.54$) than in the optical after the break
($\alpha_2=1.35$).
Thus, we conclude that the presence of the cooling break
between optical and X-ray at 80 minutes after the burst is
more likely than no break.

Although the content of some elements, as measured by optical spectroscopy,
is a matter of ongoing debate (Penprase et al.~\cite{Penprase06};
Prochaska~\cite{Prochaska05,Prochaska06a}), we use the X-ray absorption to
infer either the overabundance of at least some of the $\alpha$
elements, like oxygen, or the presence of a significant amount of
molecular gas along the line of sight through the host galaxy.
This accounts for the high X-ray absorption and the amount of
Zn measured in the optical as well as the measured dust depletion pattern.

We provided marginal evidence for the peculiarity of the
circumburst environment of GRB~051111.
The possible existence of a link between the class of the
so-called FRED GRBs and the properties of the afterglow is supported
by recent studies (Hakkila \& Giblin~\cite{Hakkila06};
Bosnjak et al.~\cite{Bosnjak06}), according
to which these bursts are more likely to be associated
with SNe. In light of this possibility, a thorough study of
the properties of this class of GRBs could help shed light
on this possible association.

\begin{acknowledgements}
We thank the referee for useful comments.
We thank S. Savaglio for reading this manuscript as well as for
useful discussions and for and R. Maiolino
for providing us with the extinction profiles.
We also thank J.X.~Prochaska for reading the manuscript and
for his comments.
We acknowledge A.~Panaitescu for useful comments.
CG thanks F. Frontera, E. Montanari and L. Amati for useful
discussions.
CG and AG acknowledge their Marie Curie Fellowships from the
European Commission.
CGM acknowledges financial support from the Royal Society.
ER and AM acknowledge financial support from PPARC.
AM acknowledges financial support of Provincia Autonoma di Trento.
MFB is supported by a PPARC Senior Fellowship.
NRT acknowledges an SRF support.
We are grateful to Scott Barthelmy and the GCN network.
The Faulkes Telescope North is operated with support from the Dill
Faulkes Educational Trust.
\end{acknowledgements}

\end{document}